\documentclass[a4paper,11pt]{article}
\pdfoutput=1 

\usepackage{jcappub} 
\usepackage[T1]{fontenc} 
\usepackage{color,soul}
\usepackage[dvipsnames]{xcolor}
\usepackage{csquotes}

\usepackage{siunitx}
\newcommand{\tsub}[1]{\ensuremath{{\scriptscriptstyle \rm #1}}}

\newcommand{\plr}[1]{\ensuremath{\left({#1} \right)}}
\newcommand{\alG}{\ensuremath{\alpha_G}}
\newcommand{\alGp}{\ensuremath{\alpha_{G'}}}
\newcommand{\suthree}{\ensuremath{SU(3)_C}}
\newcommand{\sutwo}{\ensuremath{SU(2)_L}}
\newcommand{\uone}{\ensuremath{U(1)}}
\newcommand{\uoney}{\ensuremath{U(1)_Y}}
\newcommand{\order}[1]{\ensuremath{\mathcal{O}\plr{#1}}}
\newcommand{\subl}{\ensuremath{_L}}
\newcommand{\subr}{\ensuremath{_R}}

\newcommand{\subs}{\ensuremath{_s}}
\newcommand{\subm}{\ensuremath{_\tsub{M}}}

\newcommand{\gsplit}[1]{\ensuremath{\Gamma^{\rm split}_\tsub{LPM}\plr{#1}}}
\newcommand{\bbs}{\ensuremath{\mathbb{S}}}
\graphicspath{{ArxivFigs/}}

\def\gsim{\mbox{~{\raisebox{0.4ex}{$>$}}\hspace{-1.1em}
{\raisebox{-0.6ex}{$\sim$}}~}}
\def\lsim{\mbox{~{\raisebox{0.4ex}{$<$}}\hspace{-1.1em}
{\raisebox{-0.6ex}{$\sim$}}~}}

\def\ca{C_A}
\def\cf{C_F}

\def\df{d_F}
\def\da{d_A}

\def\md{m_\tsub{th}}

\def\alphas{\alpha_{\rm s}}

\title{\boldmath Multi--Species Thermalization Cascade of
  Energetic Particles in the Early Universe}

\author{M. Drees} \author{and B. Najjari} \affiliation{Bethe Center
  for Theoretical Physics(BCTP), Bonn University,\\Nussallee 12,
  53115, Bonn Country}

\emailAdd{drees@th.physik.uni-bonn.de}
\emailAdd{bardia@th.physik.uni-bonn.de}
\abstract{Heavy long--lived particles are abundant in BSM physics and
  will, under generic circumstances, get to dominate the energy
  density of the universe. The resulting matter dominated era has to
  end before the onset of Big Bang Nucleosynthesis through the decay
  of the heavy matter component of mass $M$ into a thermal bath of
  temperature $T$. The process of thermalization primarily involves
  near--collinear splittings of energetic particles into two particles
  with lower energy. The correct treatment of these processes requires
  the inclusion of coherence effects which suppress the splitting
  rate. We write down and numerically solve the resulting coupled
  Boltzmann equations including all gauge bosons and fermions of the
  Standard Model (SM). We then comment on the dependence of the nonthermal
  spectra on the ratio $M/T$, as well as on the matter decay rate and
  branching ratios into various SM particles.}

\begin{document}
\maketitle
\flushbottom
\section{Introduction}
\label{sec:Intro}

Top--down approaches to the history of the universe, as in UV complete
theories of inflation, likely include some scalar field, e.g. the
inflaton itself or an additional \enquote{modulus} field, that is initially
displaced from its low energy minimum and later oscillates around this
minimum; if the potential around the minimum can be described by a
quadratic function this corresponds to a non--relativistic (matter)
component \cite{kt, Allahverdi:2010xz, Kane:2015jia,
  Allahverdi:2020bys}. Given enough time, a matter component will grow
to dominate the energy density of the universe because of its
different equation of state compared to radiation \cite{kt}. On the
other hand, we know from cosmological observations that the universe
was indeed radiation dominated during Big Bang Nucleosynthesis (BBN)
\cite{Giudice:2000ex, Hannestad:2004px, KaneMaybeMatter}. Hence any dominant
matter content must have decayed, and the decay products 
thermalized into the thermal bath, before the onset of BBN at a
temperature of about \SI{3}{\MeV}.

An early epoch of matter domination affects the density of
(semi--)stable relics in several ways. First of all, the entropy
released by the decay of the heavy matter particles dilutes any
possible pre--existing relic density of decoupled\footnote{We will use
  the term decoupled for both kinematic and chemical decoupling.}
species \cite{kt, Berlin:2016vnh, Berlin:2016gtr}. Moreover, for a
given temperature $T$ the Hubble expansion rate is higher than in a
radiation--dominated era, which increases the thermal freeze--out
temperature. Finally, relics can be produced non--thermally, either
directly in the decay of the heavy particles, or in collisions of
energetic daughter particles with the thermal background or with each
other \cite{Chung:1998rq, AD2, Allahverdi:2002pu, Gelmini:2006pw,
  Acharya:2008bk, Kane:2011ih, Hasenkamp:2012ii, Kurata:2012nf,
  Harigaya:2014waa, Ishiwata:2014cra, Kane:2015qea, Co:2015pka,
  Dhuria:2015xua, Kim:2016spf, Hamdan:2017psw, Drees:2017iod,
  Garcia:2018wtq, Drees:2018dsj, Chanda:2019xyl, Maldonado:2019qmp,
  Harigaya:2019tzu, Garcia:2020eof}. It has also been suggested that
the energetic decay products can produce the baryon asymmetry via
lepton-- or baryon number violating scattering reactions on the
thermal plasma \cite{Hamada:2015xva, Asaka:2019ocw}. A quantitative
treatment of production rates due to the scattering of highly
energetic decay products requires a good understanding of their number
density and energy distributions. In particular, knowledge of the
chemical composition of the thermalizing decay products is crucial.

The process of thermalization leads to the growth of the number
density of high energy particles while simultaneously reducing the
average energy of the out of equilibrium states. In cosmology, it is
well established \cite{Davidson:2000er, Allahverdi:2002pu,
  Harigaya:2013vwa, Harigaya:2014waa, Harigaya:2019tzu, Drees:2021lbm}
that the process of thermalization for gauge interacting particles is
driven by $2 \rightarrow 3$ processes; in leading log approximation,
these can be described as quasi--elastic $2 \rightarrow 2$ reactions
followed by a \emph{splitting} process, with the \emph{parent} high
energy particle breaking up to two lower energy, nearly collinear
\emph{daughter particles}. As first pointed out in
\cite{Harigaya:2013vwa}, the high collinearity of the daughter states
calls for the inclusion of coherent multiple scatterings in the
plasma, known as the Landau--Pomeranchuk--Migdal (LPM) effect
\cite{Landau:1953um, Migdal:1956tc, Baier:1998kq}. Destructive
interference between these interactions of the parent and daughter
particles with the plasma leads to a parametrically suppressed
splitting rate.

The process of energy loss of energetic gauge interacting particles in
the presence of a thermal background has also been extensively studied
in the context of Heavy Ion Collisions (HIC), and the physics of the
resulting Quark-Gluon Plasma (QGP) \cite{Arnold:2002zm, Jeon:2003gi,
  Arnold:2008zu, Kurkela:2011ti, AbraaoYork:2014hbk}. Here the
thermalization procedure evolves the initial spectrum of energetic
partons traveling mostly along the beam directions towards a thermal
distribution of quarks and gluons. Since color exchange is the
dominant interaction in the QGP, most studies focus on colored matter;
however, the emission of photons from the QGP and from electromagnetic
plasmas have also been studied using the same machinery, as the photon
spectrum serves as a probe of the QGP \cite{Arnold:2001ba,
  Arnold:2001ms, Arnold:2002ja}.

Although potentially on a very different energy scale, the physics of
thermalization following matter decay is expected to be realized via
the very same mechanisms as in the case of quark gluon plasmas
resulting from HIC experiments \cite{Berges:2020fwq}. In recent
studies in the context of thermalization in cosmology
\cite{Harigaya:2014waa, Harigaya:2019tzu}, analytical approximations
to the resulting spectrum of thermalizing particles of a single
species have been obtained. In \cite{Drees:2021lbm}, a more accurate
numerical approach has been used to find this spectrum. Since the
exchange and emission of non--Abelian gauge bosons is crucial in the
thermalization, this approach is adequate for the description of pure
Yang--Mills theories. Here the single species is a non--Abelian gauge
boson and the LPM effect can be described by a single power law
suppression. We will refer to such a treatment as the \emph{pure gauge
  treatment}.

A pure gauge approximation is motivated by the fact that gauge boson
emissions in a plasma are typically favored both by larger
$\mathcal{O}\plr{1}$ group factors and by soft gauge boson emission
enhancements, and in some cases by statistical enhancement and
blocking factors \cite{Kurkela:2018oqw}. In fact, the dominance of the
$\suthree$ interactions means that irrespective of the initial decay
products, the energy density of the plasma can be expected to flow
towards the colored sector and in particular to gluons, so that
eventually a pure gluon plasma is a better approximation than that of
$SU(2)$ gauge bosons. The single species approximation, however, does
not answer the question of how quickly the \emph{chemical}
composition\footnote{\label{foot:anti}We will always assume vanishing
  chemical potential in this work; by the chemical composition we
  merely mean the relative abundance of various species. Consequently,
  we will not distinguish between particles and antiparticles.} of the
plasma flows towards the colored QCD sector, nor is it capable of
estimating other subdominant contributions to the total out of
equilibrium number density. Note that the primary matter decay
products need not carry color. Moreover, many scattering reactions of
interest may require particles other than gluons in the initial
state. In these cases the spectra of other particles (than gluons)
become particularly important.

In this study, we extend previous works by including all non--scalar
species of the standard model in the thermalization cascade of
energetic particles. This requires including the correct form for
different vacuum splitting kernels, as well as the variations in the
LPM suppression factor corresponding to the different Abelian and
non--Abelian interactions of the charged particles involved in the
splitting process. We also systematically consider different parent
particles. We find that for sufficiently large ratios of the energy of
the parent particle to the ambient temperature the spectrum of
non--thermal particles will indeed eventually be dominated by gluons;
however, for color singlet parent particles this will happen only at
energies two to three orders of magnitude below that of the original
particle. We also observe that the relative abundances of different
 species can vary by orders of magnitude, starkly different from the 
 case of chemical equilibrium.

The remainder of this paper is organized as follows: in
sec.~\ref{sec:splitting} we review the framework of thermalization via
collinear splitting processes including the LPM
effect. Section~\ref{sec:boltzmann} includes a formulation of the
Boltzmann equations governing the number density of the various
species. Numerical results are presented in sec.~\ref{sec:solution},
whereas sec.~\ref{sec:application} briefly sketches how to compute the
production of massive particles in annihilation reactions involving
one particle from the thermal bath and one particle from the
non--thermal spectrum, followed by a summary and some concluding
remarks in sec.~\ref{sec:conclusion}.

\section{Thermalization via splitting cascades}
\label{sec:splitting}

The process of thermalization of energetic decay products in cosmology
involves an energetic particle losing energy by successive splittings
to many lower energy particles; eventually, it injects energy into the
thermal plasma. For parent particles charged under the gauge
group of the Standard Model (SM), the underlying physics is similar to
that of thermalization in heavy ion collisions. We will therefore use
the results and conventions from this field. We will, however, only be
interested in scenarios where an isotropic thermal plasma is already
in place.\footnote{In scenarios where there is no preexisting thermal
  radiation bath, e.g. at the onset of reheating after inflation, the
  energetic decay products can only interact with each other. The
  critical process will then be the formation of a \emph{seed} of soft
  particles, on which later energetic decay products can scatter
  \cite{Davidson:2000er, Allahverdi:2002pu}. This is somewhat similar
  to the initial stages of HIC \cite{Baier:2000sb,
    Kurkela:2011ti}. The details of this process, and in particular
  the critical timescale of seed formation, set the \emph{maximal}
  temperature of the radiation bath \cite{Allahverdi:2002pu,
    Harigaya:2013vwa}. We will not be dealing with these very early
  stages of thermalization. Note that in possible post--inflationary
  epochs, where energetic particles are injected, the existence of a
  thermal bath is guaranteed.} We will first introduce the basic
processes involved in the splitting cascade and the physics of the LPM
suppression, before presenting the equations used to calculate the
relevant rates.

\begin{figure}[th] \label{fig:2to2}
\centering
\includegraphics[width=0.8 \textwidth]{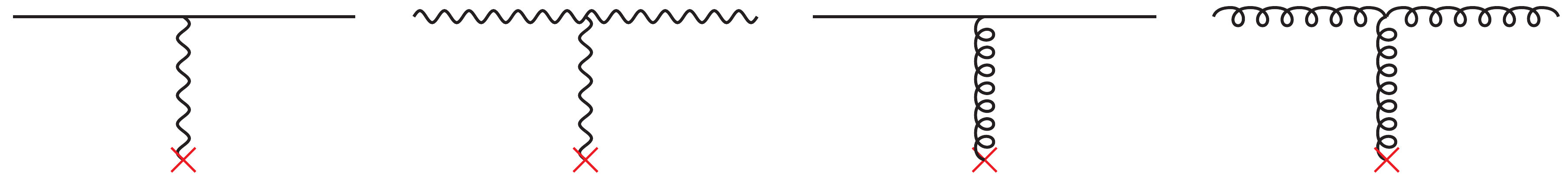}
\caption{Elastic scattering processes of an energetic particle on the
  thermal bath mediated by a gauge boson; the former is denoted by the
  line on top while the the latter is represented by the cross at the
  bottom. The intermediate gauge boson should be considered as having
  acquired the thermal mass \eqref{eq:effmass}.}
\end{figure}

Assuming the existence of a thermal bath of temperature $T$, we will
be focusing on the thermalization of energetic parent particles of
species $s$ and energy $E \approx p \gg T$ via interactions with the
bath. It has been noted that the large energy suppresses the cross
section for $2\rightarrow 2$ interactions except for $t-$channel
forward scattering via exchange of a massless gauge boson
\cite{Davidson:2000er}.\footnote{$u-$channel diagrams should be
  included in processes with identical particles in the initial
  state. See table 2 in \cite{Arnold:2002zm} for a list of
    contributing matrix elements in a gauge-mediated $2 \to 2$
    process.} The forward scattering cross section in vacuum is
infrared (IR) divergent; this divergence is regulated by the thermal
mass
\begin{equation} \label{eq:effmass}
m^\tsub{th}_G\plr{T} \approx g_G T\,,
\end{equation}
where $G \in\left\lbrace C = \suthree,W=\sutwo,Y=\uoney \right\rbrace$
denotes the dominant gauge group in the scattering process. Figure
\ref{fig:2to2} shows schematically such an elastic process, with time
flowing from left to right and the cross denoting a coupling
to a particle in the thermal bath.

The crucial point is that strict forward scattering does not reduce
the energy of the parent particle at all. In the presence of a thermal
bath most elastic scattering reactions will have
$|t| \sim \left(m^\tsub{th}_G \right)^2$, and hence typical
momentum exchange
\begin{equation} \label{eq:elasticdelk}
\delta k^\tsub{el}_G = m^\tsub{th}_G\,.
\end{equation}
The rate for these reactions is given by
\begin{equation} \label{eq:elasticrate}
  \Gamma_\tsub{el}^s \approx \tilde{g}_*^s \alpha_G T \,
   \equiv 1/t_\tsub{el}^s \,,
\end{equation}
where the factor $\tilde g_*^s$ includes the order unity group factors
for the elastic scattering of species $s$, and $t_\tsub{el}^s$ is the
the timescale between successive elastic scatterings of such a
particle.

By themselves, these soft elastic processes make for inefficient
energy transfer between the parent particle and the thermal medium,
since complete thermalization would need
${\cal O}\left( p/m^\tsub{th}_G \right)$ scatterings, i.e. the
thermalization time would scale as $1/p$. This is also true for hard
elastic scattering reactions, which can lead to much larger energy
transfer but whose rate is suppressed by $1/p$.

\begin{figure}[ht]
\centering
\includegraphics[width=0.55 \textwidth]{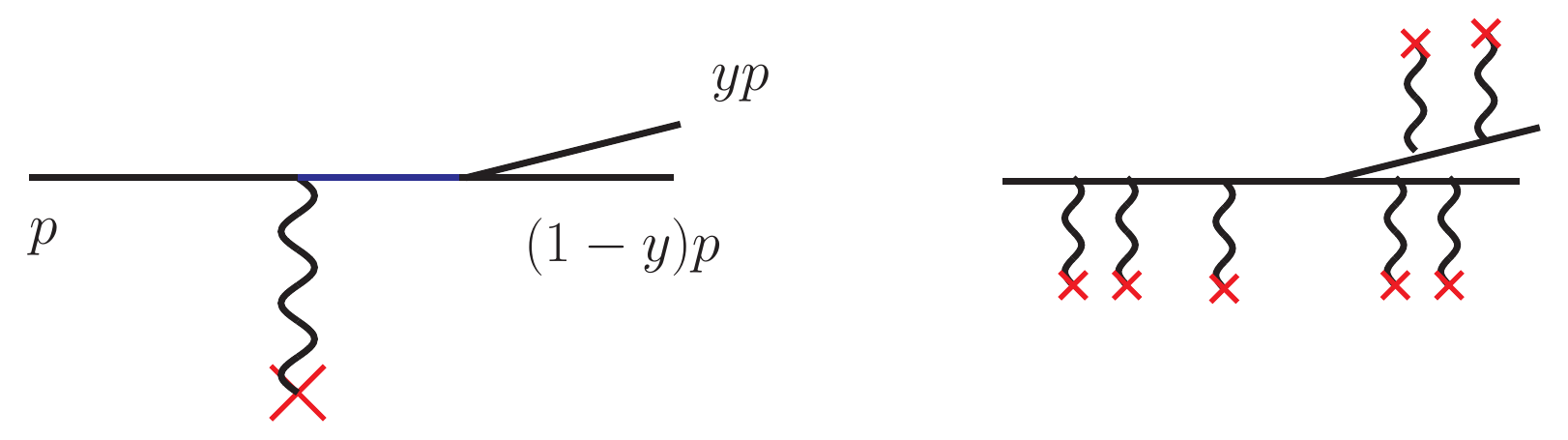}
\caption{(Left) Schematic of splitting processes after scattering on
  the thermal bath mediated by a gauge boson. Here the solid lines
  represent fermions or gauge bosons. The parent particle can shed a
  large fraction $y$ of its momentum $p$, see eq.(\ref{eq:ydef}). The
  small momentum transferred by the intermediate boson implies that
  the connecting propagator -- depicted in blue -- is close to the
  mass shell. (Right) Multiple couplings of the particles involved in
  the splitting process to the thermal bath of temperature $T$; many
  soft processes are required for the loss of collinearity and
  consequently coherence of the splitting process.}
\label{fig:2to3generic}
\end{figure}

However, scattering off the thermal plasma can kick the scattering
particle off the mass shell; the outgoing particle can then lose up to
half its energy in a $1 \rightarrow 2$ splitting
process. Fig.~\ref{fig:2to3generic} (left) shows such a splitting
reaction, and defines our convention for the momenta of the participating
particles. In contrast to elastic processes, the maximal energy loss
is not restricted by the virtuality of either the exchanged gauge
boson or of the scattered particle. The daughter particle can
therefore carry away a large fraction
\begin{equation} \label{eq:ydef}
y \equiv k/p
\end{equation}
of the parent's energy, while the virtuality of the intermediate gauge
boson is still of the order of its thermal mass given in
eq.(\ref{eq:effmass}). The corresponding small momentum transfer given
by eq.(\ref{eq:elasticdelk}) implies that the virtuality of the
scattered particle is also small. Note that the emission of the
daughter can in principle proceed via a different interaction, with a
coupling $\alpha_\tsub{G'} \leq \alpha_\tsub{G}$ than that of the
dominant $t-$channel gauge boson exchange.

The most frequent splitting processes are those involving the largest
coupling. Their rate is suppressed by a factor $\alpha_G$ relative to
that for elastic scattering. These higher order processes nevertheless
dominate the energy loss since only ${\cal O}\left(\log(p/T)\right)$
of these reactions are required for the thermalization of the parent
particle.

Were we to ignore medium effects, we would get the so--called
Bethe--Heitler differential rate for these processes:
\begin{equation} \label{eq:bethe}
  \frac{d \Gamma_\tsub{split}^\tsub{BH}}{d \log{k}} \sim \alpha_{G'}
  \Gamma_\tsub{el}^s \leq \alpha_G^2 T\,,
\end{equation}
corresponding to a time between subsequent splitting reactions
$t_\tsub{BH} \equiv
{\Gamma_\tsub{split}^\tsub{BH}}^\tsub{-1}$. However, in the rest frame
of the thermal medium, the formation time for the splitting process is
of order $p/(m_{\rm G}^{\rm th})^2$, which is much longer than the
time between successive soft elastic scatterings of
eq.(\ref{eq:elasticrate}) \cite{Arnold:2002zm}. One should therefore
coherently add up all possible contributions to a splitting process,
schematically depicted in the right fig.~\ref{fig:2to3generic}. The
medium leads to a reduced interaction rate; this is the topic of the
next subsection.

\subsection{Physics of the LPM suppression}
\label{subsec:LPMphysics}

The formal inclusion of the LPM effect requires coherently summing the
many ladder diagrams of figure \ref{fig:2to3generic}, corresponding to
extra interaction vertices; this may also be reformulated as a
variational problem that is more readily numerically tractable
\cite{Arnold:2002ja,Arnold:2001ba}. Alternatively, one may use
physical arguments to deduce the parametric form and
order--of--magnitude estimate of the LPM suppression for generic
processes. In this section, we will keep the species-dependence of the
various rates and momentum transfers explicit.

The LPM suppression results from the destructive interference among
multiple splitting matrix elements, or equivalently, from the fact
that the process is unable to physically distinguish multiple
scattering centers in the medium so long as they lie within the
coherence region of one another, thereby reducing the effective target
density as perceived by the scattering process
\cite{Arnold:2002zm}. The coherence can be understood to last as long
as the phase factor accumulated in successive soft scattering
reactions is small.

In general, both the emitting and the emitted particle undergo soft
scattering reactions on the thermal background. We will work in the
rest frame of the thermal bath, which in our application corresponds
to the cosmological rest frame. We use coordinates where the parent
particle propagates along the $z$ axis, and the splitting occurs at
$x^\mu_{\rm split} = 0$. Denoting the time elapsed traversing the
thermal bath by $\delta t$, the trajectory of the emitting particle
and the $3-$momentum of a daughter particle of energy $k \gg T$ are
then given by
\begin{equation} \label{eq:fourx}
  \vec{x}(\delta t) \simeq \delta_t (\sin\theta\, \hat{e}_\bot
  + \cos\theta\, \hat{e}_z) \,, \quad
  \vec{k} \simeq (k_\parallel \, \hat{e}_z + k_\bot \hat{e}'_{\bot})\,,
\end{equation}
where $\hat{e}_\bot$ and $\hat{e}'_\bot$ are (in general independent)
unit vectors in the $(x,y)$ plane. Recall that we are interested in
collinear splitting. The transverse momentum $k_\bot \ll k$ of the
daughter as well as the deviation $\theta \ll 1$ of the emitter from
the initial $z$ direction therefore only result from the soft
interactions shown in fig.~\ref{fig:2to3generic}, so that we can
approximate
$\sin\theta \simeq \theta, \, \cos\theta \simeq 1 - \theta^2/2$ and
$k_\parallel \simeq k - k_\bot^2/(2k)$.

Crudely, one may say that the coherence, and hence the interference,
persists as long as the accumulated phase satisfies
\begin{equation} \label{eq:phase1}
  \delta \phi = k \cdot x \simeq \delta t \left( - k_\bot \theta \hat{e}_\bot
    \cdot \hat{e}'_\bot + k \frac{\theta^2}{2}
    + \frac {k_\bot^2}{2 k} \right)  \leq 1\,.
\end{equation}
While this condition is satisfied, further splittings are not possible.
The angle $\theta\plr{\delta t}$ is given by
\begin{equation} \label{eq:theta}
\theta(\delta t) \simeq p_\bot(\delta t)/p\,.
\end{equation}
It is also time dependent. Altogether the accumulated phase is thus
of order
\begin{equation} \label{eq:phase2}
  \delta \phi \sim \delta t \left( \frac{k_\bot p_\bot}{p} \oplus
    \frac {k p_\bot^2}{2p^2} \oplus \frac{k_\bot^2}{2k} \right)\,.
\end{equation}
The $\oplus$ symbol indicates that the terms should be added in
quadrature in a statistical sense, since in general the emitter and
both emitted particles undergo a random walk through multiple scatters
on the thermal background. During a time span $\delta t$ there will be
$\delta t / \delta t_{\rm el}$ \enquote{steps}, with \enquote{step size}
$\delta k_{\rm el}$. If both particles interact with similar strength,
we thus have
\begin{equation} \label{eq:krandomwalk}
  \left\langle k_\bot \right\rangle (\delta t)
  \simeq \left\langle p_\bot \right\rangle 
  \simeq \sqrt{ \frac {\delta t} {\delta t_\tsub{el}} } \, \delta k_\tsub{el}
  \simeq {\delta t}^{1/2} \Gamma_\tsub{el}^{1/2}
  \delta k_\tsub{el}\,.
\end{equation}
For very hard splitting, where $k \sim p$, all three terms in
eq.(\ref{eq:phase2}) will then be of the same order. However, in the
more likely configuration where $k \ll p$ the third term
dominates. For the case where all particles that participate in the
splitting have roughly equal coupling strength to the thermal bath, we
can thus approximate the phase $\delta \phi \sim \delta t k_\bot^2/k$,
with the understanding that $k$ denotes the momentum of the {\em
  softer} daughter particle. The coherence time, defined by $\delta \phi
(t_{\rm coh}) = 1$, is then given by
\begin{equation} \label{eq:Sec1:cohtime}
  \delta t_\tsub{coh} \sim \sqrt{ \frac{k} {\Gamma_\tsub{el}
      \delta k_\tsub{el}^2} }\,,
\end{equation}
where we have used eq.(\ref{eq:krandomwalk}). Recall that the next
splitting reaction can only happen at $t > \delta t_\tsub{coh}$; this
can be described by reducing the plasma density and therefore the
interaction rate by a factor
\begin{equation} \label{eq:rLPM}
  R_\tsub{LPM}(k) \simeq \delta t_\tsub{ coh}^{-1} \delta t _\tsub{el}
  =\sqrt{\frac{\delta k_\tsub{el}^2}{k \Gamma_\tsub{el} }}\,.
\end{equation}
Please note that this suppression factor scales like $1/\sqrt{k}$, which
favors softer splittings due to their shorter coherence time.

Recall, however, that we have assumed equal interaction strengths for
all three particles participating in the splitting process. Let us now
consider the case where the emitted particle basically does not couple
to the thermal bath, as in $q\rightarrow \gamma q$. In the particular
case of an Abelian gauge boson, the absence of $t-$channel soft
scattering processes on the thermal bath implies that its
momentum can be considered a constant (originating from the splitting
process itself). We can then choose its direction, rather than that of the
parent particle, to define the $z$ axis, i.e. $k_\bot = 0$. In this case,
only the second term in eq.(\ref{eq:phase2}) survives. Again defining
the coherence time via $\delta \phi(\delta t_{\rm coh}) = 1$ and using
eq.(\ref{eq:krandomwalk}) for $p_\bot$ we now have
\begin{equation} \label{eq:abelrLPM}
  \delta t_\tsub{coh} \sim \sqrt{\frac{p^2}
    {k \Gamma_\tsub{el} (\delta k_\tsub{el})^2}}, \quad
  R_\tsub{LPM}(k) \sim \sqrt{\frac{k \delta (k_\tsub{el})^2}
    {p^\tsub{2} \Gamma_\tsub{el} }}.
\end{equation}
Evidently, the LPM effect favors the emission of {\em harder} photons,
in contrast to what we had in \eqref{eq:rLPM} for gluon emission.
This stark difference can be understood from the observation that a
harder photon implies a softer second daughter particle, whose soft
elastic scatterings with the plasma can therefore result in a faster
growth of $\theta$ and hence earlier loss of coherence.

Before we move on to presenting explicit expressions for the
LPM--corrected rates, calculated in the literature, and for the many
available processes of the standard model fermions and gauge bosons,
we may summarize the results of our physical arguments for the LPM
suppressed splitting rate as follows. The rate for a process can be
decomposed as the result of the splitting rate in vacuum, using the
thermal gauge boson mass (\ref{eq:effmass}) as infrared regulator, and
the corresponding LPM suppression factor:
\begin{equation} \label{eq:dressedrate}
  \frac {d\Gamma^{\rm split}_\tsub{LPM}\left(s(p)\rightarrow s'(k) + s''(p-k)
  \right)} {d k} =
  \frac {d\Gamma^{\rm split}_\tsub{vac}\plr{s(p)\rightarrow s'(k)+ s''(p-k)}}
  {d k} \times R_\tsub{LPM} \plr{k,s'}\,,
\end{equation}
where 
\begin{equation} \label{eq:rLPMsummary}
  R_\tsub{LPM} \plr{k} \propto \left\lbrace
 \begin{aligned}
        & \sqrt{ \frac{ k T}{p^2}}  & & s' = \texttt{Abelian GB}& \\
        & \sqrt{ \frac {T} {\min{\plr{k,p-k}} }} & & \texttt{others} & 
\end{aligned}
       \right..
\end{equation}
Here $s$ is the parent particle and $s', \, s''$ are the two daughter
particles.

Of course, the heuristic derivation given here does not allow to
derive exact numerical factors. In particular, we have assumed that
the transverse momentum due to the splitting itself is no larger than
$\delta k_{\rm el}$. The inclusion of processes with momentum transfer
$q > \delta k_{\rm el}$ enhances the final splitting rate by a
so--called Coulomb logarithm of order $\ln(p/T)$. Once this factor is
included, we will see in the next subsection that the results of careful
calculations indeed approximately reproduce the behavior given by
eqs.\eqref{eq:dressedrate} and \eqref{eq:rLPMsummary}.

\subsection{LPM suppressed splitting rates in leading logarithmic approximation}
\label{subsec:leading}

As mentioned earlier, the study of relativistic heavy ion collisions
requires knowledge of various elastic $2 \rightarrow 2$ scattering as
well as splitting processes, where the inclusion of coherence effects
is crucial. We mostly rely on the results obtained in
\cite{Arnold:2008zu}, whose notation we follow; we, however,
generalize the results to allow for two separate gauge groups being
responsible for the splitting process and the gauge-mediated
scattering of the three particles of the thermal bath. We also discuss
the proper choice of parameters corresponding to each of these two
gauge groups. Note that we are interested in the splitting rate per
parent particle, while in the literature the total emission of
daughter particles per time and volume is often given
\cite{Arnold:2002ja, Arnold:2001ba, Arnold:2001ms}. Moreover, we
neglect all Bose enhancement and Fermi blocking factors; this is
justified since we are only interested in particles with energy much
above $T$, which have very low occupation numbers.

To be precise, the LPM corrected rate for the various splitting
reactions can be written as
\begin {equation} \label{eq:dGamma}
\frac{d\Gamma^\tsub{split}_\tsub{LPM}\plr{s(p)\rightarrow s'(k) + s''(p-k)}}{dy}
  = \frac{(2\pi)^3}{p \nu_s^G} \,
    \gamma_{s \rightarrow s's''}\plr{p, yp, (1-y)p} \,.
\end{equation}
Here $y$ is the momentum fraction carried by the species $s'$ as in
eq.(\ref{eq:ydef}), and $\nu_s^G$ is the number of spin degrees of
freedom for the species $s$ times $d^\tsub{G}\subs$, the dimension of
its gauge representation under the gauge group $G$. For example, a
gluon has $\nu_g^{SU(3)} = 2 \times 8 = 16$, and a quark has
$\nu_q^{SU(3)} = 6, \, \nu_q^{SU(2)} = 4$.\footnote{Once electroweak
  interactions are included one has to distinguish between fermions of
  different chirality, since in the SM only left--handed fermions and
  right--handed antifermions have $SU(2)$ interactions. We will
  comment in the next subsection on the required changes to our
  expressions.}

The bulk of information about the splitting process is encoded the in
the \emph{splitting functions}
$\gamma_{s \rightarrow s' s''} \plr{p; yp, (1-y)p}$. Here $s, s'$ and
$s''$ stand for either a fermion $F$ or a gauge boson $A$. To leading
logarithmic approximation, they are given by
\cite{Arnold:2008zu}\footnote{As mentioned before, results from the
  literature and in particular \cite{Arnold:2008zu} deal with the case
  $G=G'=\suthree$.}:
\begin{subequations} \label{eq:gammas}
\begin {align}
 \gamma_{A\rightarrow AA}(p; yp, (1-y)p)
   & = \frac{d_A^{G'} C_A^{G'} \alGp}{(2\pi)^4 \sqrt2}
   \frac{1+y^4+(1-y)^4}{y^2(1-y)^2} &\cdot &  
 \left[ \md^2 \, \hat\mu_\perp^2(1,y,1{-}y;A,A,A) \right]_G\,;
\label{eq:gamma_ggg}
\\
   \gamma_{F\rightarrow AF}(p; yp, (1-y)p)
   &= \frac{d_F^{G'} C_F^{G'} \alGp}{(2\pi)^4 \sqrt2}
    \frac{1+(1-y)^2}{y^2(1-y)} &\cdot &
 \left[  \, \md^2 \, \hat\mu_\perp^2(1,y,1{-}y;F,A,F) \right]_G\,;
\label{eq:gamma_qgq}
\\
   \gamma_{A\rightarrow FF}(p; yp, (1-y)p)
   &= \frac{d_F^{G'} C_F^{G'} \alGp}{(2\pi)^4 \sqrt2}
    \frac{y^2+(1-y)^2}{y(1-y)} \times N_{fl} &\cdot &
\left[ \md^2 \, \hat\mu_\perp^2(1,y,1{-}y;A,F,F)\right]_G\,.
\label{eq:gamma_gqq}
\end{align}
\end{subequations}
Here we have again labeled with $G$ the gauge group responsible for
the (dominant) scattering processes on the thermal background, while
gauge group $G'$ is responsible for the splitting. In our numerical
analysis we will ignore subdominant contributions to the soft
scattering; e.g. we will only include gluon exchange for the soft
scattering of quarks. In contrast, the gauge group $G'$ is determined
uniquely by the identity of the involved gauge boson(s). In detail,
eqs.(\ref{eq:gamma_ggg}), (\ref{eq:gamma_qgq}) and (\ref{eq:gamma_gqq})
describe pure gauge splittings like $g \to gg$ or $W \to WW$, gauge
boson emission from a fermion like $q \to gq$ or $l\to Wl$, and
splitting of a gauge boson to a fermion antifermion pair like
$g \to q\bar q$ or $W \to l \bar l$, respectively; of course,
$\gamma_{A \rightarrow A A} = 0$ for $U(1)$ interactions.

The first term in each splitting function captures the physics of the
splitting process without the plasma effects, and thus loosely
corresponds to $\Gamma_\tsub{vac}$ in \eqref{eq:dressedrate}. The
representation dimension $d\subs^\tsub{G'}$ was introduced below
equation (\ref{eq:dGamma}), and the remaining parameter
$C\subs^\tsub{G'}$ is the quadratic Casimir. Within the SM, with the
fermions and bosons in fundamental and adjoint representations
respectively, we have for the group $SU(N)$:
\begin {equation} \label{eq:sungroupfacs}
   \cf = \plr{N^2-1}/2N \,,
   \qquad
   \ca =  N \,,
   \qquad
   \df = N \,,
   \qquad
   \da = N^2-1,
 \end {equation}
while for a $U(1)$, we have
\begin {equation} \label{eq:uonegroupfacs}
   \cf^\tsub{s} = Y\subs^2,
   \qquad
   \ca = 0 \,,
   \qquad
   \df = 1 ,
   \qquad
   \da = 1 \,,
 \end {equation}
with $Y\subs$ the $U(1)$ charge of the corresponding fermion.

The $y$ dependence of the different fragmentation functions is
determined by the structure of the relevant three--point vertex; up to
a factor $1/[y(1-y)]$ it is the same as that of the splitting
functions appearing in the famous DGLAP equations \cite{Gribov:1972ri,
  Dokshitzer:1977sg, Altarelli:1977zs}. The parameter $N_{fl}$ in
eq.(\ref{eq:gamma_gqq}) quantifies the number of indistinguishable Dirac
fermions that the gauge boson $A$ can split into. For the $SU(2)$ and
$U(1)_Y$ gauge bosons splitting into $q \bar q$ pairs this will always
include a color factor of $3$; $N_{fl}$ may be larger if several
quarks (of different generations, say) are considered
indistinguishable.

\begin{figure}[ht]
\centering
\includegraphics[width=0.3 \textwidth]{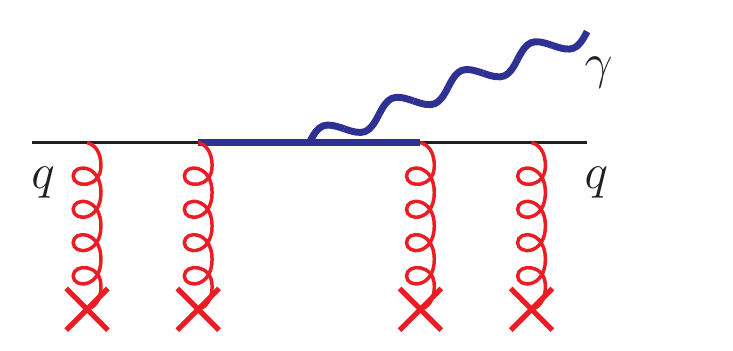}
\caption{Photon bremsstrahlung (blue, $G'=\uone_\tsub{EM}$) from a quark
  interacting strongly with colored plasma particles
  (red, $G=\suthree$). The resulting rate is of order
  $\alpha_G \alpha_{G'}$.}
\label{fig:q2gammaq}
\end{figure}

The second sets of terms in eqs.(\ref{eq:gammas}), denoted by the
subscript $\left[\ldots \right]_G$, can be thought of as representing
the background plasma of temperature $T$, describing the scattering
target density, the momentum transfer for the $2\rightarrow 3$
scattering process, as well as the coherent suppression
effects. Before moving on to the individual elements, let us take a
moment to clarify the assignment of $G$. Take as an example the
physical process of emission of a photon from a QGP via the splitting
process $q \to \gamma q$ depicted in figure \ref{fig:q2gammaq}. In the
previous part, we assigned $G'=U(1)_\tsub{EM}$ as the gauge group
responsible for the emission of the photon, corresponding to the blue
sections of figure \ref{fig:q2gammaq}. This splitting process however
relies on the interactions with the hot plasma; the same interactions
also provide the coherent suppression effect, shown in red in figure
\ref{fig:q2gammaq}. As can be seen, the latter class of processes will
be dominated by the strong interactions of the colored quarks, so that
$G=\suthree$ but with $C_\gamma = 0$. Note that we ignore electroweak
interactions of the quarks with the background plasma in this
treatment; similarly, we ignore hypercharge interactions of $SU(2)$
doublet leptons with the background. In this approximation only a
single group $G$ contributes to the loss of coherence.\footnote{Recall
  that this formalism resums a large number of scattering
  diagrams. Since diagrams where different gauge bosons are exchanged
  can in general interfere in these reactions, the contribution of
  electroweak interactions to the scattering of quarks cannot simply
  be treated by incoherently summing over several group factors in
  eqs.(\ref{eq:gammas}).}

The exact leading order expression for the thermal mass of the gauge
boson, whose order of magnitude has been given in
eq.(\ref{eq:effmass}), is \cite{Arnold:2008zu}
\begin{equation} \label{eq:mth_ex}
  m^2_{G,{\rm th}} = \frac{1}{3} g_G^2 T^2 \left( C_A^G + \sum_f \frac{d_f}{d_A} N_f C_F^G
  \right)\,,
\end{equation}
where $N_f$ is the total number of Dirac fermions with a gauge
representation corresponding to $C_F$; see eqs.(\ref{eq:sungroupfacs})
and (\ref{eq:uonegroupfacs}) for groups $SU(N)$ and $U(1)$,
respectively. In these terms, the function $\hat\mu_\perp^2$ on the
right--hand side (RHS) of eqs.(\ref{eq:gammas}) can in leading log
approximation be universally written as \cite{Arnold:2008zu}
\begin{align} \label{eq:muLL}
&
  \hat\mu_\perp^2(y_1,y_2,y_3;s_1,s_2,s_3)
  \simeq \frac{g_GT}{m_{G,{\rm th}}} \left[\frac{2}{\pi} \, y_1 y_2 y_3 \,
    \frac{p}{T} \right]^{1/2}
  \Biggl\{ \Biggl[
       \tfrac12 (C_{s_2}+C_{s_3}-C_{s_1}) y_1^2
       + \tfrac12 (C_{s_3}+C_{s_1}-C_{s_2}) y_2^2
\nonumber\\ & \hspace{15em}
       + \tfrac12 (C_{s_1}+C_{s_2}-C_{s_3}) y_3^2
    \Biggr] \ln(\sqrt{p/T})  \Biggr\}^{1/2}\,.
\end{align}
Equation \eqref{eq:muLL} encodes the LPM effect for a splitting
process $s_1 \rightarrow s_2 s_3$; the various Casimir factors
represent the coupling of the three particles involved in the thermal
plasma to the thermal bath. The coupling constant $g_G$ cancels in
eq.(\ref{eq:muLL}) due to $m_{G,{\rm th}}$ appearing in the
denominator, see eq.(\ref{eq:mth_ex}).

We may now see how the explicit form of eq.(\ref{eq:muLL}) reproduces
our physically motivated results for both cases in eqs.(\ref{eq:rLPM})
and \eqref{eq:abelrLPM}. Consider, first, the photon emission process
discussed in section \ref{subsec:LPMphysics}. As emphasized
above, the photon has $C_{s2}=C_A=0$, while $C_{s1}=C_F=C_{s3}=4/3$
for the colored quarks, so that
\begin{equation} \label{eq:muLLphoton}
  \hat\mu_\perp^2(1,y,1-y;q,\gamma,q)
  \propto  \left[ \frac{2}{\pi} \, y (1-y) \, \frac{p}{T}  \right]^{1/2}
  \Biggl\{ \frac{4}{3} y^2 \ln(\sqrt{p/T})  \Biggr\}^{1/2}\,.
\end{equation}
On the other hand, for gluon emission from a quark, $q \to gq$, we
have $C_{s2}=C_A=3$, and $C_{s1}=C_F=C_{s3}=4/3$, so that
\begin{equation} \label{eq:muLLgluon}
  \hat\mu_\perp^2(1,y,1-y;q,g,q) \propto \left[ \frac{2}{\pi} \,
    y (1-y) \, \frac{p}{T}  \right]^{1/2} 
  \Biggl\{\Biggl[\frac{4}{3} y^2 + 3 \left(1+(1-y)^2\right)
        \Biggr]  \ln(\sqrt{p/T}) \Biggr\}^{1/2}\,.
\end{equation}
For small $y$, where the term in the second square parentheses in
eq.(\ref{eq:muLLgluon}) approaches a constant,
eqs.(\ref{eq:muLLphoton}) and \eqref{eq:muLLgluon} differ by factor
$y=k/p$, reproducing the relation between eqs.(\ref{eq:rLPM}) and
\eqref{eq:abelrLPM}.\footnote{The various group and $\order{1}$
  factors, e.g. in eqs.(\ref{eq:gammas}) and \eqref{eq:muLL}, had been
  subsumed into the constant $\sqrt{\tilde{g_*}}$ in
  ref.\cite{Drees:2021lbm} and in eq.(\ref{eq:elasticrate}).} Finally,
the appearance of the Coulomb logarithm is, as mentioned before, the
result of including all process with a momentum transfer larger than
$m_\tsub{th}$, which occur at a rate smaller than that of 
eq.(\ref{eq:elasticrate}).

Before concluding this subsection we comment on processes where two
participants carry color but the third one only undergoes weaker
non--Abelian interactions. Within the SM, the only such processes
involve two $SU(2)$ doublet quarks and an $SU(2)$ gauge boson. An
example is $q \to Wq$ splitting, which is described by
eq.(\ref{eq:gamma_qgq}) with $G'=SU(2),\, G=SU(3)_C$. Recall that we
simply set $\ca=0$ when treating the emission of a $U(1)$ gauge boson,
which cannot scatter on the background by $t-$channel exchange of a
gauge boson. On the other hand, the non--Abelian $W$ will have such
interactions. Of course, the relevant $SU(2)$ coupling strength is
significantly smaller than that of $SU(3)$. Nevertheless, the
discussion at the end of sec.~\ref{subsec:LPMphysics} showed that the
scattering of the $W$ might terminate coherence if it has much less
energy than the parent quark. We treat this by assigning
$\ca=(\alpha_W/\alpha_S) \ca^W$ to the $W$ boson in the splitting
process. This may be a crude approximation, but we do not know of a
study focusing on the calculation of the LPM effect involving multiple
gauge groups. Moreover, as we will see in sec.~\ref{sec:boltzmann},
the energy loss of the original parent particle, and hence the
spectrum of all daughter particles, is mostly determined by more
symmetric splittings, where the difference between $C_A = 0$ and
$0 < C_A \ll 1$ is small.

\subsection{Particle content and the treatment of chirality}
\label{subsec:particles}

Equations \eqref{eq:gammas} and \eqref{eq:muLL} provide the rate of
all splitting processes involving SM fermions and gauge bosons. Before
we move on to composing the system of Boltzmann equations, however, we
should address some technicalities regarding the applicability of the
splitting rates to various processes.

The formalism we adopt here has been developed for QCD interactions,
which are vector--like. As already noted, the standard model being a
chiral theory, we will keep track of the chiralities of the fermions
in the splitting cascade.

Since we always sum or average over the spins of the participating
gauge bosons, we do not need to change the $y$ dependence of any
splitting function. Moreover, the $1/2$ factors in the normalization
of $\nu$ in eq.(\ref{eq:dGamma}) and of $d_F$ in
eq.(\ref{eq:gamma_qgq}) cancel each other so that no change is
required for the case of gauge boson emission off a chiral fermion.
However, since $N_{fl}$ in eq.(\ref{eq:gamma_gqq}) counts the number
of indistinguishable Dirac fermions, it should include a factor of $1/2$
when considering species of chiral (or Weyl) fermions.

Note that the gauge boson vertices are chirality conserving. The
relaxation of any preexisting net\footnote{As we will see in
  sec.~\ref{sec:solution}, the relative chiral asymmetry is strongly
  diluted by the fact that the thermalization cascades increases,
  manifold, the number of fermion pairs, a majority of which will be
  the result of chiral symmetric $\suthree$ and $\uone$ gauge boson
  splittings.} chiral charge via $2\rightarrow 2$ or $1\rightarrow 2$
processes will therefore rely on bare fermion mass insertions
\cite{Boyarsky:2020cyk, Boyarsky:2020ani}, which are absent in the
unbroken $SU(2)$-phase, or involve emission or exchange of a scalar
Higgs boson;\footnote{If we wanted to include the Higgs doublet $\phi$
  among the parent or daughter particles, we would need to include
  several additional splitting functions for processes involving
  spin$-0$ bosons. Since $\phi$ does not have strong interactions, and
  describes a rather small number of degrees of freedom, we ignore
  these processes in our treatment. This should be a good
  approximation, unless Higgs bosons feature prominently among the
  original parent particles.}  the latter processes are highly
suppressed by the small Yukawa couplings of most SM fermions, 
as well as by a factor of $y^2/2$ in the splitting
  function relative to that for the emission of a gauge boson
  \cite{Reya:1979zk, Abramowicz:1982jd}. In what follows, we will
  therefore treat the chiralities of fermions as conserved quantities,
  i.e. we will present separate spectra for left-- and right--chiral
  fermions $f_{L/R}$.

Somewhat trivially, one does not need to distinguish between the
members of a given representation of the gauge group. As an example,
in the phase of unbroken $SU(2)$ we would not, and in fact cannot,
distinguish between an up--type and a down--type left--chiral
fermion. Even if we fix the gauge (so that \enquote{up--type} is well
defined), $2 \to 2$ scattering processes on the thermal background
induce changes between these states at the high rate
$\Gamma_\tsub{el}$, i.e. these states are not well--defined over the
duration of a splitting process. The same is also true for the color
of a quark, so that the quark undergoing a splitting process in figure
\ref{fig:2to3generic} cannot be considered to have a well--defined
color even in a fixed gauge.

However, at least in principle different right--chiral fermions $f_R$
are physically distinct particles. For example, $u_R$ and $d_R$ have
different $\uoney$ charges. However, as it is difficult to imagine
physical scenarios where we would need to distinguish between the up
and down type $f_R$, we will simply treat both these species as
possessing an average squared charge in \eqref{eq:uonegroupfacs},
using
\begin{equation} \label{eq:Yeffsq}
Y_{q_R}^2 = \frac12 \plr{Y^2_{u_R}+Y^2_{d_R}}\,.
\end{equation}

Moreover, we will not distinguish between fermions of different generations
This still leaves us with seven distinct particle species:
\begin{equation} \label{eq:particlecontent}
  s\in \left\lbrace q_L,\,q_R,\,\ell_L,\,\ell_R,\,g,\,W,\,B\right\rbrace
  \equiv \bbs.
\end{equation}
Note also that we assume equal production of particles and
antiparticles. This is true if the original $M$ particles always
decay into $s \bar s$ pairs. Treating more fermion species as
distinguishable is in principle straightforward. However, this would
increase the number of coupled Boltzmann equations that need to be
solved, and also the number of terms in some of these equations. The
choice \eqref{eq:particlecontent} should be sufficient to illustrate
the main effect of the appearance of particles with very different
interaction strengths in this first, exploratory analysis.

\section{System of Boltzmann equations}
\label{sec:boltzmann}

We established in section \ref{sec:splitting} that $2 \to 3$ processes
play a significant role in the redistribution of energy, and in the
growth of number density, towards a thermal distribution. These can be
understood as quasi--elastic scattering processes that leave the
energy $p$ of the parent particle almost unchanged, followed by
$1 \to 2$ splitting which distributes the energy of the parent among
two daughter particles of energy $k$ and $p-k$, respectively. The
rate of this $2 \rightarrow 3$ process, differential in the energy
of one of the daughters, is given by eq.(\ref{eq:dGamma}).

Our final goal is to find the number densities of all particles that
are not in thermal equilibrium during the thermalization cascade.
These densities are governed by a set of Boltzmann equations. In order
to introduce the framework and our notation, it might be beneficial to
first briefly review the treatment of a pure gauge theory where one
only has to consider a single species of particle, as done in
\cite{Drees:2021lbm, Harigaya:2014waa}.

\subsection{Pure gauge cascade}
\label{subsec:pureg}

The time dependence of the phase space distribution of a particle
species is, as usual, given by a Boltzmann equation. The well--motivated assumptions of homogeneity and isotropy imply that the phase
space or number density depends solely on the magnitude $p$ of the
$3-$momentum and on the cosmological time $t$. Splittings of the
the species $s$ generate a non--thermal spectrum, which we denote by
\begin{equation} \label{eq:Sec1-spectrum}
  \tilde{n}\subs\plr{p} \equiv \frac{dn\subs(p)}{dp} \quad \texttt{such that}
  \quad \int_{T}^{p_{\rm max}} \tilde{n}\subs(p) dp = n\subs\,.
\end{equation}
The generic Boltzmann equation governing this number density reads
\begin{equation} \label{eq:boltzmann1}
  \frac{\partial}{\partial t} \tilde{n}\subs( p, t )
  - 3H p \frac{\partial}{\partial p} \tilde{n}\subs ( p, t )
  = +\mathcal{C}\tsub{inj}(p,t) - \mathcal{C}\tsub{dep}(p,t).
\end{equation}
The two collision terms on the RHS represent injection processes
adding particles of momentum $p$ and depletion processes removing
particles of momentum $p$, respectively. As a typical primary injection
process, we consider the decay of a long--lived matter particle of mass
$M$, so that $p_{\rm max} = M/2$. The injection also includes the
\enquote{feed-down} from particles with momentum $k > p$ through their
splittings. In general, the form of the initial injection spectrum
will be model dependent. Without loss of generality, however, we may
assume the matter particle to decay to two initial particles, so that
the initial injection can be written as $\delta-$function contribution
at $p = M/2$. Since the Boltzmann equation is linear in $\tilde{n}_s$,
the result for any other injection spectrum can be obtained by 
convoluting this initial spectrum with the final spectrum of non--thermal
particles resulting from the initial delta function.

It can be shown that in almost all scenarios of interest the initial
particles thermalize within a time which is much shorter than a Hubble
time \cite{Harigaya:2013vwa,Drees:2021lbm}. This means that the
term $\propto H$ in eq.(\ref{eq:boltzmann1}) can be neglected;
furthermore, the temperature of the thermal bath can be
considered as constant within the thermalization time. Injection and
depletion will then quickly reach equilibrium, as long as new particles
with initial energy $M/2$ keep being injected. This (quasi) steady--state solution satisfies
\begin{equation} \label{eq:boltzmann2}
  \mathcal{C}\tsub{dep}(p) = \mathcal{C}\tsub{inj}(p).
\end{equation}
The exchange symmetry of the two daughter states for the case of
$g \to gg$ splittings, see eq.(\ref{eq:gamma_ggg}), allows one to
write eq.(\ref{eq:boltzmann2}) as\footnote{In \cite{Drees:2021lbm} an
  approximation of eq.(\ref{eq:gamma_ggg}) was used, where the
  $y-$dependence was approximated as $1/y^{3/2}$ for $y \leq 1/2$, and
  the region $y > 1/2$ was treated via the $y \leftrightarrow 1-y$
  exchange symmetry. Since this approximates the full result to better
  than $10\%$ for $y < 0.4$ (and hence for $y > 0.6$ as well), this
  does not change the solution qualitatively.}
\begin{align} \label{eq:boltzmann3}
 2n_\tsub{ M} \Gamma_\tsub{ M} \delta(p-M/2)
 &+ \int_{p + \kappa T}^{M/2} \tilde{n}\subs(k)
   \frac{d \Gamma^{\rm split}_\tsub{ LPM}(s(k) \to s(p)s(k-p))}{d p} d k
   \nonumber \\
 &=  \int_{\kappa T}^{p/2} \tilde{n}\subs (p)
 \frac{d \Gamma^{\rm split}_\tsub{ LPM} (s(p) \to s(k)s(p-k))}{d k} d k \,.
\end{align}
Here $n\subm$ and $\Gamma\subm$ denote the number density and decay
rate of the long--lived matter particle, and the splitting rates on
both LHS and RHS are given by eqs.(\ref{eq:dGamma}) and
\eqref{eq:gamma_ggg}. We have used $\kappa T$ as an infrared (IR)
cutoff for the $2 \rightarrow 3$ processes, where $\kappa$ is of order
unity. This is reasonable since the much faster $2 \rightarrow 2$
scattering reactions allow momentum exchange of order $T$; moreover,
the total spectrum of particles with energy $\leq 3T$ is in any case
dominated by the thermal contribution. It has been shown
\cite{Drees:2021lbm} that the precise choice of $\kappa$ does not
affect the physical solution for $p \gg T$, and so we may choose
$\kappa=1$ for convenience. The choice of an IR cutoff further allows
for a straightforward numerical solution of eq.(\ref{eq:boltzmann3}).

This numerical solution is facilitated by introducing
\begin{equation} \label{eq:dimensionless}
\widetilde{N}\subm = \frac {2 n\subm \Gamma\subm} 
{\Gamma^{\rm split}_\tsub{ LPM}(M/2)} \,, \quad
\bar{n}\subs\plr{p} = \frac{\tilde{n}\subs\plr{p}}{\tilde{N}\subm},
\end{equation}
and further using the dimensionless quantities
\begin{equation} \label{eq:xdef}
x = \frac{p}{T}\, \implies \bar{n}\subs\plr{x} = T \bar{n}\subs\plr{p}\,.
\end{equation}
If the coupling is treated as a constant, the spectrum $\bar{n}(x)$ can
only depend on the single parameter
\begin{equation} \label{eq:xm}
x\subm = \frac{M}{2T}\,.
\end{equation}

The Boltzmann equation \eqref{eq:boltzmann3} can now be written as
\begin{equation} \label{eq:boltzmannX}
\bar{n}\subs(x;x\subm) = \int_{x + \kappa}^{x\subm} 
\frac {\bar{n}\subs(x')} {\gsplit{s(x) \to ss}}
\frac {d \gsplit{s(x') \to s(x)s(x'-x)}} {dx} dx'
  + \delta(x-x\subm)\,.
\end{equation}
Here $\gsplit{s(x) \to ss}$ is the total (integrated) rate for a
particle of dimensionless momentum $x$ to undergo a splitting
process. Note that the splitting rate appears in both the numerator
and denominator of eq.(\ref{eq:boltzmannX}), so that numerical factors
like $\sqrt{\tilde{g_*}^s}$ as well as the coupling strength $\alG$ do
not affect $\bar{n}\subs\plr{x}$; they appear in the final, physical
spectrum only via the normalization factor $\tilde{N}\subm$.

The numerical solution of eq.(\ref{eq:boltzmannX}) can be approximated
by \cite{Drees:2021lbm}
\begin{equation} \label{eq:nbarsingle}
  \bar n\subs(x;x_\tsub{ M}) = \delta(x-x\subm) +
  \frac {\left[ a \left( x/x_\tsub{ M} \right)^{-3/2}
      \left( 1 - x/x_\tsub{ M} \right)^{-b}  + c \right]
  \left( 1 - 2/\sqrt{x_\tsub{ M}} \right) }
{ \sqrt{x_\tsub{ M}}   \left( 1 - \sqrt{2/x} \right)^{5/4} }\,,
\end{equation}
with $a,\, b,\, c \approx 1/2$. The form of the single species
solution \eqref{eq:nbarsingle} shows that for $x\subm > x \gg 1$, the
solution is proportional to the function
\begin{equation}\label{eq:ffunction}
f\plr{x/x\subm} \equiv \left[ a \left( x/x_\tsub{ M} \right)^{-3/2}
      \left( 1 - x/x_\tsub{ M} \right)^{-b}  + c \right]\,,
\end{equation}
which depends only on the ratio $x/x\subm$.  In section
\ref{sec:solution} we will use this latter feature, and the
pure--gauge solution to present the results of the splitting cascade
involving all species $s \in \bbs$.

\subsection{Multiple species and the coupled set of integral equations}

So far we have assumed that the parent particle of species $s$ can
only split into two daughter particles of the same species, as is the
case for a pure Yang--Mills gauge theory. The splitting of a gluon
into two gluons is indeed the fastest splitting reaction in the SM,
and is expected to dominate the evolution of the system produced in
heavy ion collisions, at least at central rapidities
\cite{Kurkela:2018oqw}. However, the multiplicity of accessible
fermion flavors suggests that a sizable population of fermionic
daughters develops in the splitting cascade. Since SM quarks are
charged under several gauge groups, eventually electroweak gauge
bosons, and hence leptons, will become part of the cascade even when
starting from a parent gluon. Moreover, the matter particles whose
decays feed the cascade might primarily decay to colorless species.
Recall finally that the purpose of computing the spectrum of particles
in the cascade triggered by the primary decay is to compute rates for
processes that leave observable relics, e.g. dark matter particles or
a baryon asymmetry; and these reactions may involve preferentially, or
even only, colorless particles in the initial state. Hence a treatment
including all SM species is necessary.

We will therefore extend the previous studies by formulating and
solving the coupled system of Boltzmann equations governing the energy
spectrum of all particles listed in eq.(\ref{eq:particlecontent})
during a LPM--suppressed thermalization cascade. As in 
subsec.~\ref{subsec:pureg} we will be interested in the quasi steady--state
solution. The spectrum of particles of species $s$ is then determined by
the integral equation
\begin{eqnarray} \label{eq:Sec2-BoltzmannCoupled}
 2n_\tsub{ M} \Gamma_\tsub{ M} Br\subs \cdot \delta(p-M/2)
  & + & \sum_{s',s''} \int_{p + \kappa T}^{M/2} \tilde{n}_\tsub{s'}(k)
        \frac{d \Gamma^{\rm split}_\tsub{ LPM}
   (s'(k) \rightarrow s(p) s''(k-p))}{d p} d k \nonumber \\
  & = & \sum_{s',s''} \int_{\kappa T}^{p-\kappa T} \tilde{n}\subs (p)
        \frac{d \Gamma^{\rm split}_\tsub{LPM} (s(p) \rightarrow s'(k) s''(p-k))}
        {d k} d k \,.
\end{eqnarray}
Here $2Br\subs$ is the average number of $s$ particles produced in a
given decay of an $M$ particle. As noted earlier, we assume equal
production of fermions and antifermions, and the density $n_s$ in fact
includes $s$ and $\bar s$ particles if these are distinct.\footnote{If
  only decays of the type $M \rightarrow s \bar s$ decays are allowed,
  $Br\subs$ is the branching ratio for one such mode. However, if
  $M \rightarrow s s'$ with $s \neq s'$ decays are possible, the
  corresponding branching ratio would appear as $2 Br\subs$ in the
  equation for $\tilde n_s$, and as $2 Br_{s'}$ in the equation for
  $s'$.} In the rate
$\Gamma^\tsub{split}_\tsub{LPM} \plr{s(p)\rightarrow s'(k) s''(p-k)}$
the particle type of $s''$ is often fixed once $s$ and $s'$ have been
specified; this is true in particular if $s$ and/or $s'$ denotes a
gauge boson. However, if $s$ and $s'$ are both fermions (which implies
$s=s'$), several types of gauge boson might be possible for $s''$; for
$q_L \rightarrow q_L$ splitting, $\sum_{s''}$ will even run over all
three species of gauge bosons of the SM.

In eq.(\ref{eq:Sec2-BoltzmannCoupled}) most dependence on high--scale
physics has been factored into the first term. In particular, the
product $n_{\rm M} \Gamma_{\rm M}$ only affects the overall
normalization of all spectra. In contrast, the splitting rates can be
computed within the theory that is valid at energy scales well below
$M$, which we assume to be the SM. The second and third term in
eq.(\ref{eq:Sec2-BoltzmannCoupled}), which determine the shapes of the
various spectra and also affect their relative normalization,
therefore depend on high--scale physics only through the upper limit
of integration of the first integral. As in the case of the single
particle cascade, it is useful to make these dependencies explicit by
using \eqref{eq:xdef} to rewrite \eqref{eq:Sec2-BoltzmannCoupled} in
the dimensionless form
\begin{eqnarray} \label{eq:coupleddims}
\bar{n}_s(x) &=& \frac {\Gamma^{\rm split}_\tsub{LPM}\plr{s^*,x\subm}}
                {\Gamma^{\rm split}_\tsub{LPM}\plr{s,x\subm}}
                Br_s \cdot \delta\plr{x-x\subm} \nonumber \\
             &+& \sum_{s',s''} \int_{x+\kappa}^{x\subm}
                 \frac {\bar{n}_\tsub{s'}\plr{x'}}
                 {\Gamma^{\rm split}_\tsub{LPM}\plr{s,x}}
                 \frac{d \Gamma^{\rm split}_\tsub{LPM}\plr{s'(x')
                    \rightarrow s(x) s''(x'-x)}} {dx} dx' \,.
\end{eqnarray}
Here we have generalized eq.(\ref{eq:dimensionless}) by normalizing
the spectra of all species using the total rate of $2 \rightarrow 3$
reactions of some reference species $s^*$:
\begin{equation} \label{eq:multinormalization}
\widetilde{N}^*\subm = \frac{2 n\subm \Gamma\subm}
{\Gamma^{\rm split}_\tsub{LPM}(s^*,M/2)} \,, \quad
\bar{n}_s(x) = T \bar{n}_s\plr{p} =
T \frac {\tilde{n}\subs\plr{p}} {\tilde{N}\subm^*}\,,
\end{equation}
with
\begin{equation} \label{eq:Sec2-dimlesstotalrate}
  \Gamma^{\rm split}_\tsub{LPM}(s,x) = \sum_{s' ,s''} \int_\kappa^{x-\kappa}
 \frac {d \Gamma^{\rm split}_\tsub{LPM} \plr{s(x) \rightarrow s'(x') s''(x-x')}}
{dx'} dx'\,.
\end{equation}
We use the particle with the largest total rate, i.e. the gluon
$s^*=g$, for the normalization. 

In a single particle cascade, increasing the coupling strength
increases the total rate for $2 \rightarrow 3$ reactions which {\em
  reduces} the overall normalization of the spectrum. The reason is
that a larger rate of splitting reactions decreases the thermalization
time, i.e. the particle spends less time in the cascade. In the case
at hand, this is true only if all couplings are increased by the same
factor. In contrast, increasing some coupling relative to that of the
reference particle (i.e.  relative to the QCD coupling, for our choice
$s^* = g$) still reduces the normalization of the first term in
eq.(\ref{eq:coupleddims}), but it also {\em increases} the probability
that $s$ will be produced later in the cascade.
\section{Numerical calculation of the spectra}
\label{sec:solution}

The numerical solution of eq.(\ref{eq:coupleddims}) proceeds
essentially in the same way as in the case of a single particle cascade:
one starts at $x = x_\tsub{M}$ (regularizing the $\delta$ function
\cite{Drees:2021lbm}) and successively works down to lower $x$. Of
course, now we actually need to solve seven such equations, for the
species listed in eq.(\ref{eq:particlecontent}).

\begin{figure}[h!]
\centering
\includegraphics[width=1 \textwidth]{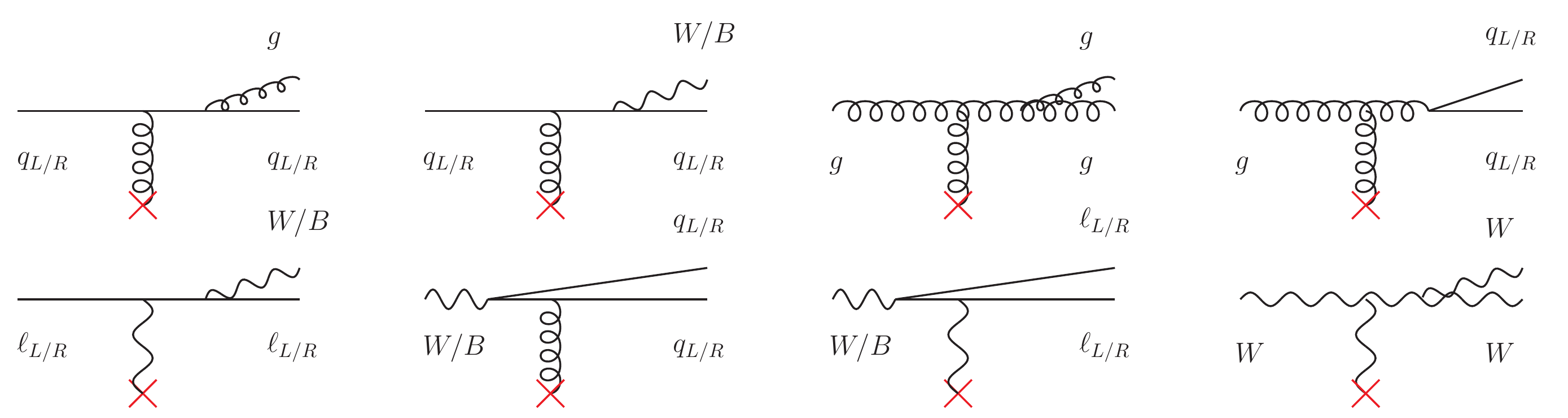}
\caption{Splitting processes for the species in $\bbs$ that can
  proceed via soft interactions with the thermal background. The red
  crosses again denote couplings to particles in the thermal bath
  via many soft exchanges. We only include the $t-$channel gauge boson
  which has the largest possible coupling, as discussed in section
  \ref{subsec:leading}; e.g. the process $B \to \ell\subl \ell\subl$
  is mediated by an $SU(2)$ gauge boson $W$. Note also that $W$ bosons
only couple to left--handed fermions $q_L, \, \ell_L$.}
\label{fig:splittings}
\end{figure}

The integrand in eq.(\ref{eq:coupleddims}) is computed from
eqs.(\ref{eq:dGamma}) and \eqref{eq:gammas}, using
eqs.(\ref{eq:sungroupfacs}) and \eqref{eq:uonegroupfacs} for the
coupling parameters; we use three generations of massless leptons and
quarks, and include extra factors of $1/2$ for splitting of gauge
bosons to chiral fermion pairs in \eqref{eq:gammas}.  The contributing
splitting reactions are depicted in fig.~\ref{fig:splittings}. Note
that we use the parameter $x$ of eq.(\ref{eq:xdef}) in the Boltzmann
equations, while the fragmentation functions are written in terms of
the momentum fraction $y$ of eq.(\ref{eq:ydef}); the two are related
by
$y= k_\tsub{daughter}/p_\tsub{parent} =
x_\tsub{daughter}/x_\tsub{parent}$. The use of the exact (leading
order) splitting functions \eqref{eq:gammas} implies that the total
rates $\Gamma^{\rm split}_\tsub{LPM}(s,x)$ also have to be computed
numerically, whereas the approximations used in
ref.\cite{Drees:2021lbm} allowed the integral in
eq.(\ref{eq:Sec2-dimlesstotalrate}) to be calculated analytically.

In our numerical work, we will assume that only one of the seven
species $I\in \bbs$ is produced in primary $M$ decays, i.e.
$Br_I = 1$, in which case the first term in eq.(\ref{eq:coupleddims})
is absent for the other species, $Br_s = 0\ \forall s\neq I$. We
denote the resulting scaled number densities as
$\bar{n}\subs^I\plr{x}$. Scenarios where more than one $Br_s$ is
nonzero can be treated by weighted sums of our results. If only
primary decays of the kind $M \rightarrow ss$ are
allowed\footnote{Recall that $s$ describes antiparticles as well.},
this sum reads:
\begin{equation} \label{eq:transfer}
  \bar{n}\subs\plr{x,x\subm} = \sum_{I \in \bbs} Br_\tsub{I}
  \bar{n}\subs^\tsub{I}\plr{x,x\subm}\,.
\end{equation}

Note also that the total rates
$\Gamma^{\rm split}_\tsub{LPM}(s,x_\tsub{M})$ in
eq.(\ref{eq:coupleddims}) are independent of the cosmological
parameters, so they need to be calculated only once for every species
$s$ and given value of $x\subm$.

\subsection{Solutions for $x\subm = 10^4$}

We are now ready to present some numerical results. As noted, we will
assume $Br_I=1$, but we will show results for all seven possible
choices of $I$ so that spectra for more general primary decays can be
computed using eq.(\ref{eq:transfer}). In this subsection we choose
$x\subm=10^4$; the dependence on $x\subm$ will be discussed in the
next subsection. Since all relevant vertices involve at least one
particle with virtuality of order $gT$, we use running couplings taken
at that scale. We assume $T=100$ TeV, where the $SU(2) \times U(1)_Y$
symmetry is still unbroken so that the (bare) masses of all particles
in the cascade vanish. Note that the choice of temperature affects the
couplings only logarithmically.

In the following figures spectra of particles charged under $\suthree$
($g,\, q\subl,\, q\subr$) are shown in green; spectra of colorless
particles that are charged under $\sutwo$ ($W,\, \ell\subl$) are shown
in red; and spectra of particles that have only $\uoney$ interactions
($B,\,\ell\subr$) are shown in blue. We use solid lines for gauge
bosons, while left-- and right--chiral fermions are represented by
dashed and dotted lines respectively. The black line shows the total
spectrum of particles that are not in thermal equilibrium,
$\bar{n}_\tsub{tot}\plr{x} = \sum\subs \bar{n}\subs \plr{x}$; note
that in eq.~\eqref{eq:multinormalization} $\bar{n}\subs$ are related
to the physical spectra $\tilde{n}\subs$ by a universal
factor. Finally, we show for comparison the pure gluonic (single
species) solution of eq.(\ref{eq:boltzmannX}) in dark green. As we
expect and will see, this pure gluon solution serves as an attractor
for the gluon number density, irrespective of the matter decay
branching ratios $Br_I$.

\begin{itemize}
\item \textbf{$I \in \{B, \ell_R\}$:}\\
  Let us begin with the case of an initial injection of particles that
  have only $\uoney$ interactions, $B$ and $\ell_R$. The resulting
  spectra are shown in figure \ref{fig:x4BlR}. A first observation is
  that for large $x$ the spectra of $B$ and $\ell_R$ lie well above
  the single species pure gluon spectrum. This is because the
  coefficient in front of the $\delta$ function in
  eq.(\ref{eq:coupleddims}) is considerably larger than unity here,
  due to the small total splitting rate for the colorless species as
  compared to gluons.

  Note that $B$ can split into quarks with a rate of order
  $\alpha_S \alpha_Y$ in couplings, corresponding to the vacuum and
  LPM contributions discussed in section \ref{sec:splitting}; here
  $\alpha_S$ and $\alpha_Y$ are the \enquote{fine structure constants}
  for $\suthree$ and $\uoney$, respectively. On the other hand, the
  only splitting reaction for $\ell_R$ is emission of a $B$ boson with
  rate of order $\alpha_Y^2$ in couplings. The difference in process
  rates is further enhanced by the large color, generation and flavor
  final--state multiplicity factors in the case of $B$ splittings to
  quarks.  Moreover, due to the LPM suppression factor of
  eq.(\ref{eq:muLLphoton}) the differential rate for
  $\ell_R \rightarrow B \ell_R$ splittings only scales like
  $1/\sqrt{y}$ at small $y$, just like the LPM suppressed differential
  rate for $B \rightarrow ff$ splittings does. As a result, the total
  splitting rate is considerably higher for $B$ than for $\ell_R$. The
  combination of these effects explain why at large $x$ the spectra
  for $I=B$ (left frame) lie well below those for $I=\ell_R$ (right
  frame).

  A peculiarity of the case of $I=B$ is that the first splitting of
  the original $B$ completely depletes its spectrum, trading it for a
  weighted sum of fermion spectra.\footnote{The splitting processes
    are obviously statistical in nature and so there is strictly
    speaking a left--over population of red--shifted $B$
    bosons. However, as argued above, in all relevant situations we
    have $H \ll \Gamma_\tsub{split}$, in which case the density of
    these redshifted particles is exponentially small.} Let us first
  discuss the couplings involved in these splittings. The relative
  rates of these splitting reactions are proportional to the sum of
  squared hypercharges of fermions described by a species $s$;
  including color factors, these are $5$ for $q_R$, $3$ for $\ell_R$,
  $3/2$ for $\ell_L$ and $1/2$ for $q_L$. Of course, the factor
  $m_{\rm th}^2$ in eq.(\ref{eq:gamma_gqq}) is larger for quarks by a
  factor $\alpha_\tsub{S}/\alpha_\tsub{W}$; they scatter more readily
  off the thermal background. However, this is overcompensated by the
  much larger total splitting rate of quarks, which appears in the
  denominator in eq.(\ref{eq:coupleddims}). The net effect is that the
  quark spectra in the left frame of fig.~\ref{fig:x4BlR} are
  suppressed by a factor of order $\alpha_Y/\alpha_S$ compared to
  $\ell\subr$. Note that the relative factor of $10$ larger
  hypercharge squared for right--chiral quarks together with the
  almost identical $\suthree$ dominated splittings rates of left-- and
  right--chiral quarks in the denominator of
  eq.~\eqref{eq:coupleddims} result in a larger $q\subr$ spectrum
  compared to that of $q\subl$.  Similarly, the $\ell_L$ spectrum is
  suppressed by $\alpha_Y/\alpha_W$, compared to that of $\ell\subr$,
  with $\alpha_W$ being the $\sutwo$ coupling. Finally, and most
  importantly, the differential rate for $F \rightarrow A F$
  splittings scales like $1/y^{3/2}$ if, and only if, $A$ is a
  non--Abelian gauge boson; we saw in the previous paragraph that
  otherwise it scales like $1/\sqrt{y}$. The total splitting rates for
  quarks and $\ell_L$ leptons, at a momentum scale $x$, are therefore
  enhanced by a factor $\sqrt{x}$ relative to that of $\ell_R$. This
  explains why the flux of $\ell_R$ is by far the dominant one at
  large $x$.

\begin{figure}[!ht]
\begin{minipage}{.5\linewidth}
\centering
\includegraphics[scale=0.45]{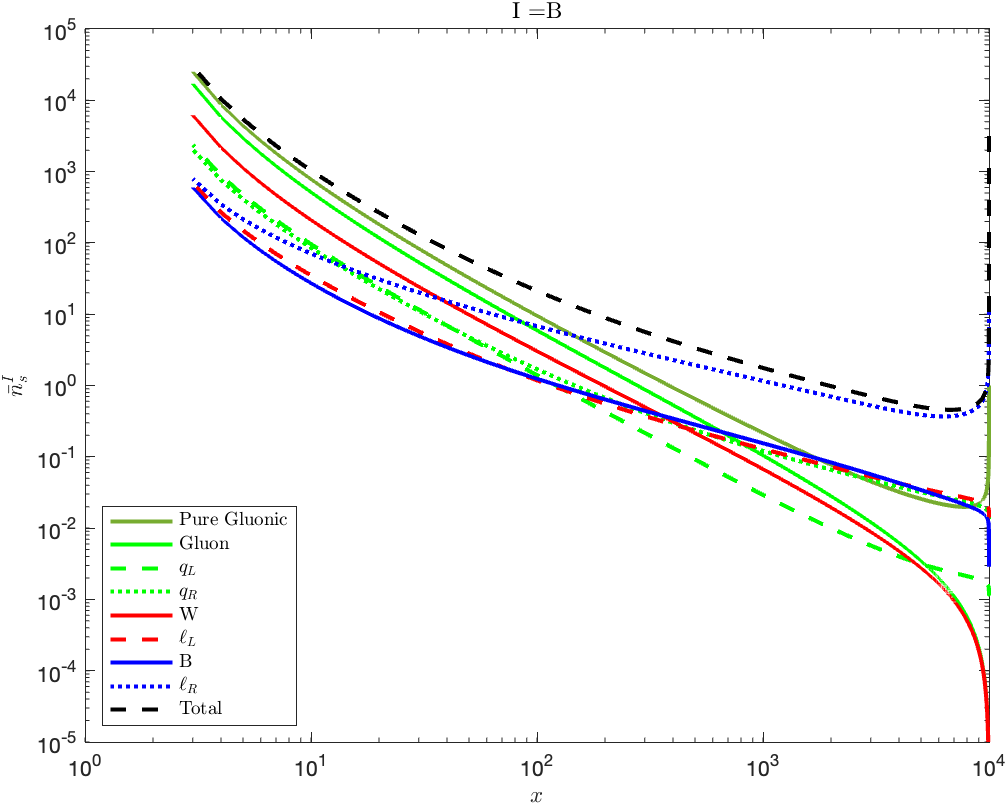}
\end{minipage}
\begin{minipage}{.5\linewidth}
\centering
\includegraphics[scale=0.45]{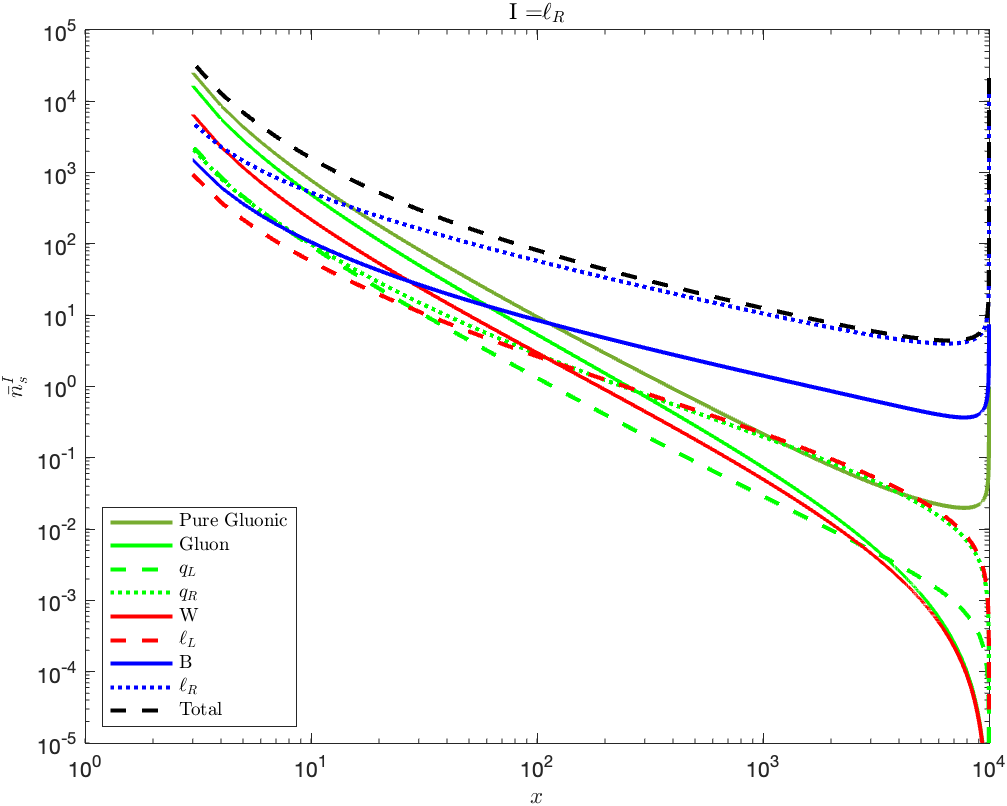}
\end{minipage}
\caption{Scaled number density functions $\bar{n}_s^I\plr{x}$ of
  eq.(\ref{eq:transfer}), for the various particles $s \in \bbs$, with
  $I=B$ (left) and $I=\ell_R$ (right); we use $x\subm=10^4$ for both
  cases.}
\label{fig:x4BlR}
\end{figure}

The right frame in figure \ref{fig:x4BlR} shows results for the case
of $\ell_R$ injection. As already noted, the single splitting channel
for the $SU(2)$ singlet leptons is emission of a $B$ boson, where the
LPM effect partially counteracts the vacuum preference for a soft
$B$. Other species of fermions are produced later in the cascade, via
splittings of the $B$, hence their relative ordering can be understood
as in the case of initial $B$ injection. The other gauge bosons can at
the earliest be produced in tertiary splitting reactions, e.g.
$\ell_R \rightarrow B \rightarrow q_L \rightarrow g,\, W$, hence their
spectra fall off fastest as $x \rightarrow x\subm$.

Despite the discussed relatively later production in the splitting cascade,
strongly interacting particles eventually dominate the non--thermal cascade.
 For original injection of $B$
bosons, where quarks and gluons are already among the products of
primary and secondary splitting reactions, this occurs for
$x/x\subm < 10^{-2}$, while for original injection of $\ell_R$ leptons
this species dominates the total flux as long as $x/x\subm >
10^{-3}$. Note that at $x \ll x\subm$ the gluon number density
approaches that of the pure gluon solution as advertised.

While the eventual dominance of particles with the strongest
interactions in the cascade seems intuitively reasonable, within our
formalism the reason for this dominance is actually rather subtle,
relying on the interplay of various charge assignments, couplings, and
group factors. Let us compare, as an example of this interplay, the
spectrum of gluons to that of $W$ bosons. In fig.~\ref{fig:x4BlR}
these spectra are very similar as $x \rightarrow x\subm$. Consider the
case of initial $B$ injection; the case of initial $\ell_R$ injection
is similar, except that one more cascade step is required to produce
$W$ or $g$, as noted above. Starting from an initial population of $B$
bosons, we saw above that the quark spectra at large $x$ are
suppressed relative to the $\ell_L$ spectrum by a factor
$\alpha_W/\alpha_S$, reflecting the ratio of the total splitting rates
of $\ell_L$ and $q$; we also mentioned that the population of $q\subr$
is an order of magnitude larger than that of $q\subl$. Nevertheless,
the LPM induced order $\alpha_\tsub{S}/\alpha_\tsub{W}$ larger
splitting rate of $q\subl \to W$ compared to $\ell_L \rightarrow W$
splittings results in both processes contributing to $W$ production
parametrically at the same order in couplings in the RHS of
eq.(\ref{eq:coupleddims}). The gluon spectrum on the other hand,
receives contributions from the quark spectra, suppressed relative to
$\ell\subl$ by a factor of $\alpha_\tsub{W}/\alpha_\tsub{S}$; combined
with the production and thermalization of gluons both being of order
$\alpha_\tsub{S}^2$ in couplings, this would result in a subdominant
gluon spectrum at high $x$, if not for the relatively large $q\subr$
population. These arguments show how the coincidence of the gluon and
$W$ spectra at high $x$ in the case of initial $B$ injection relies on
the interplay of various charge assignments, group factors, and
couplings.

In the above argument, we have assumed that the total splitting rate
of non--Abelian gauge bosons is dominated by $A \to AA$
splittings. This is in fact correct for $x \gg 1$, since this rate
scales like $1/y^{3/2}$ for $y \ll 1$. In contrast, the rate for
$A \to FF$ only scales like $1/\sqrt{y}$, hence this contribution to
the total splitting rate is suppressed by a factor $1/\sqrt{x}$
relative to that from $A \to AA$ splitting. We should emphasize that
in case of the $W$, the soft gauge boson enhancement of $W \to WW$ is
only partly compensated by a relative enhancement factor
$\alpha_S/\alpha_W$ for the $W \to q\subl q\subl$ process, reflecting
the larger scattering rate of quarks on the thermal
background. In contrast, both the weaker coupling in
  the splitting matrix element and the somewhat stronger LPM
  suppression disfavor $q_L \rightarrow q_L W$ splitting relative to
  $q \rightarrow qg$ splitting. As a result, there is more flow from
  only weakly interacting particles to strongly interacting ones than
  vice versa.

Finally, in fig.~\ref{fig:x4BlR} and the subsequent figures we have
terminated the curves at $x = 3$, i.e.  $p = 3T$, which approximates
the average energy of particles in the thermal bath. In fact, we
expect the total flux to be dominated by thermal particles out to
considerably larger energies, due to the (assumed) long lifetime of
the matter particles whose decays trigger the cascade and the short
thermalization time. Note also that the spectrum we compute becomes
independent of the IR cut--off $\kappa$ only for $x \geq 10$ (as well
as $x\subm - x \geq 10$) \cite{Drees:2021lbm}. It is worth mentioning
that this happens to be roughly the same scale at which the elastic
$2 \to 2$ processes get to compete with the inelastic processes,
further restricting the precision of results for $x \leq 10$.

\begin{figure}[!ht]
\begin{minipage}{.5\linewidth}
\centering
\includegraphics[scale=.45]{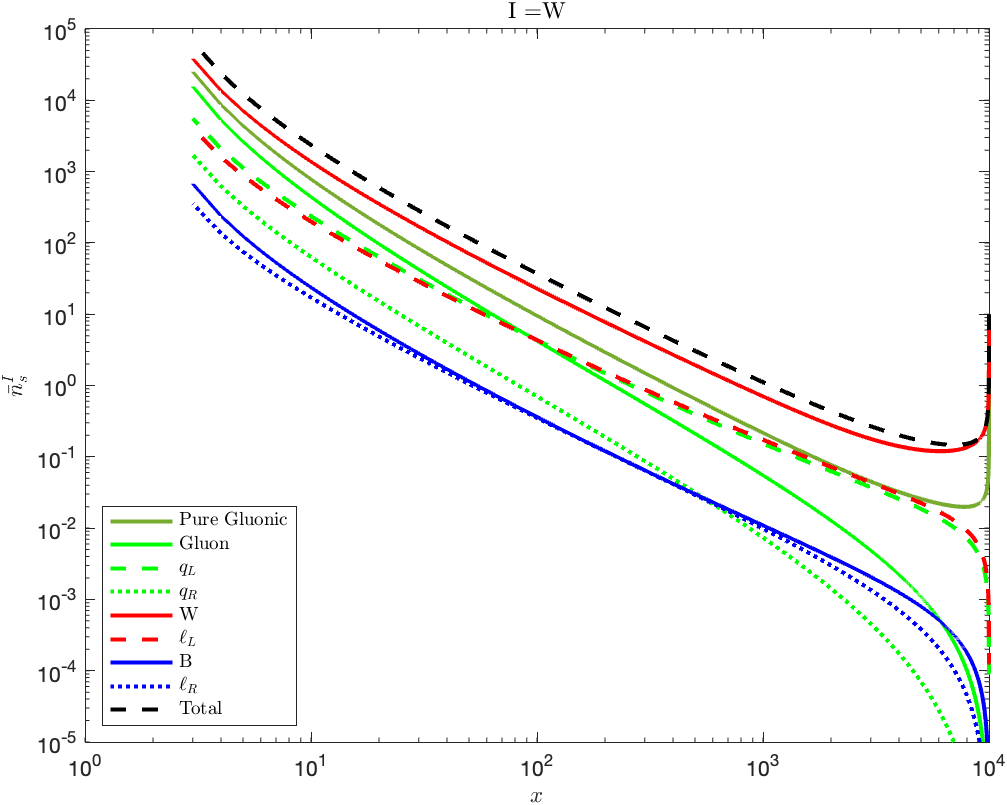}
\end{minipage}
\begin{minipage}{.5\linewidth}
\centering
\includegraphics[scale=.45]{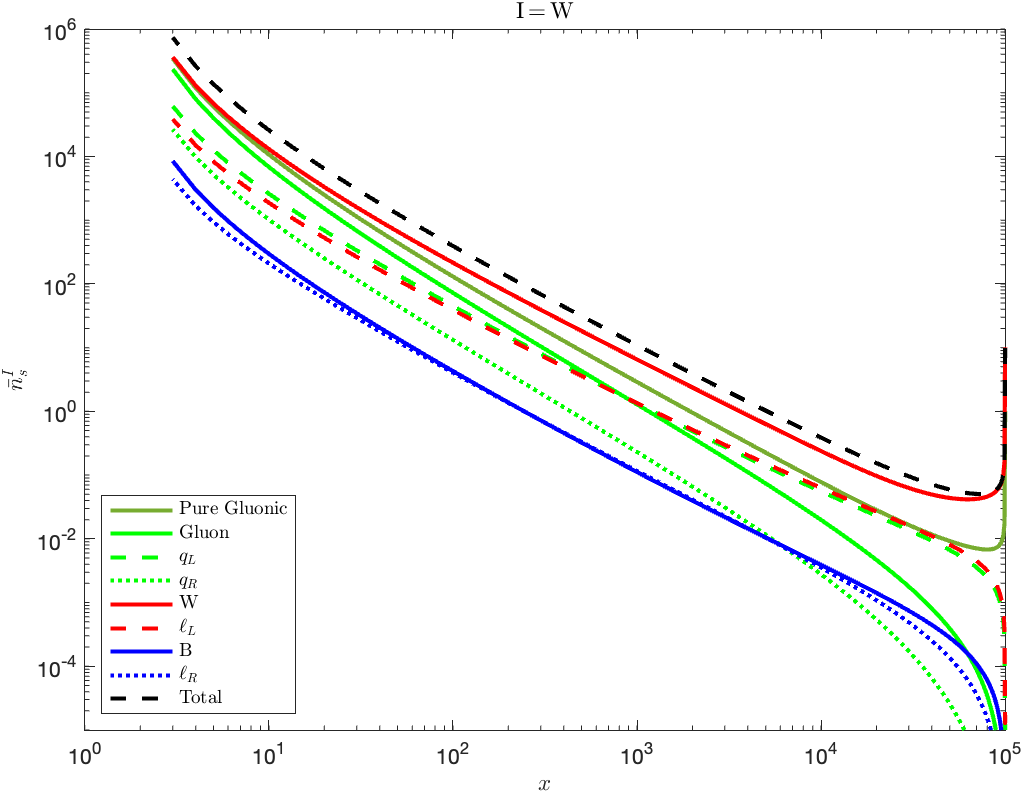}
\end{minipage}\par\medskip
\centering
\includegraphics[scale=.55]{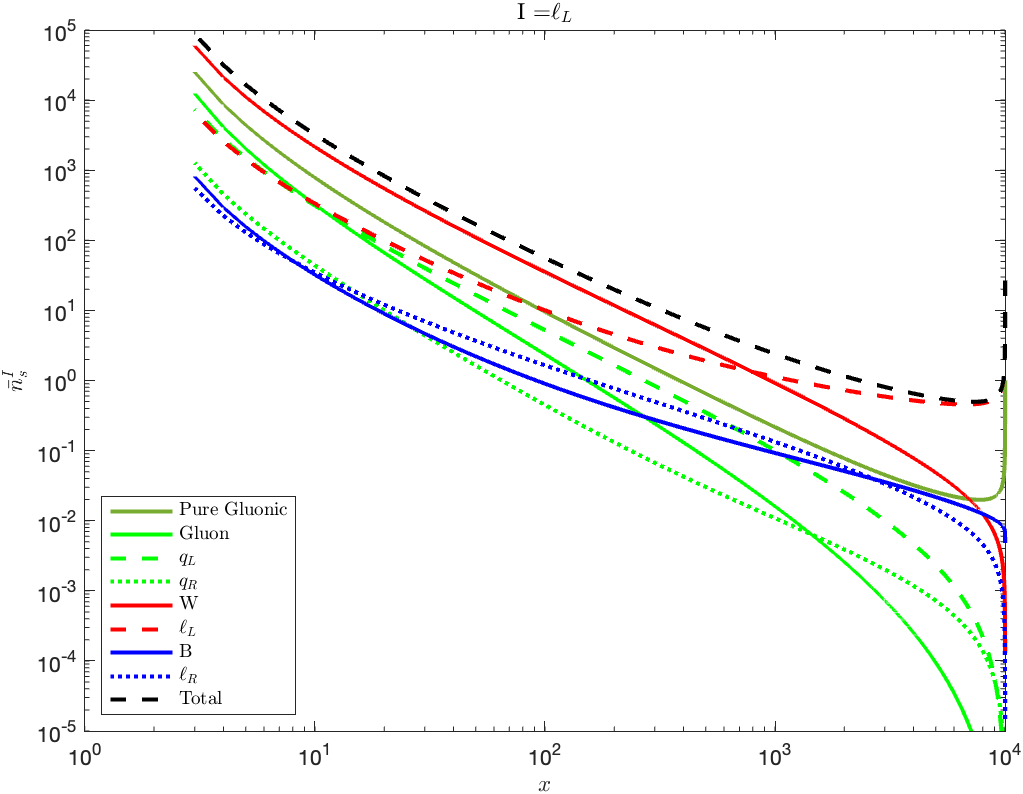}
\caption{Scaled number density functions $\bar{n}_s^I\plr{x}$ of
  eq.(\ref{eq:transfer}) for the various particles $s \in \bbs$, with
  $I=W$ (top) and $I=\ell_L$ (bottom); we use $x\subm=10^5$ in the
  top--right frame in order to show that an extra decade in $x$ allows
  the composition of thermalizing particles to approach domination by
  colored particles, in particular gluons. The top--left and bottom
  frames are for $x\subm=10^4$.}
\label{fig:x45WlL}
\end{figure}

\item \textbf{$I \in \{\ell_L,\, W\}$:}\\
  Next, we turn to the injection of color singlet particles with
  $\sutwo$ interactions, which then dominate their scattering off the
  thermal background. Results are shown in fig.~\ref{fig:x45WlL}. The
  discussion following fig.~\ref{fig:x4BlR} explains why the overall
  normalization at large $x$ lies in between that of the pure gluon
  case and the case where the injected particles have only $\uoney$
  interactions. Moreover, we again see that at sufficiently small $x$
  colored particles begin to dominate the non--thermal cascade.

  In detail, let us first consider the case of $W$ injection (top
  frames). We see that the flow towards the dominance of colored
  states is considerably slower than for the injection of $B$
  bosons. This is partly because $\sutwo$ interactions in the SM are
  stronger than $\uoney$ interactions and thus closer to the strong
  $\suthree$ processes, but mostly because $\sutwo$ is a non--Abelian
  group allowing $A \rightarrow AA$ splittings; eqs.(\ref{eq:gammas})
  show that the corresponding rate is enhanced by a factor
  $1/[y(1-y)]$ relative to that of $A \rightarrow FF$
  splittings. Therefore the original $W$ bosons strongly prefer to
  emit additional soft $W$ bosons, rather than splitting into quark
  antiquark pairs which leads to a transition to colored particles. An
  analogous argument also shows why the flux of non--Abelian gauge
  bosons eventually overtakes that of $\sutwo-$charged fermions in the
  case of fermion injection: only the relatively inefficient
  $A \rightarrow FF$ splittings increase the number of fermions in the
  cascade, while each $A \rightarrow AA$ and $F \rightarrow AF$
  splitting increases the number of gauge bosons; the rate for the
  latter is still enhanced by a factor $1/y$ relative to that for
  $A \rightarrow FF$.

  Due to the high efficiency of $W \rightarrow WW$ splitting, for
  $x\subm = 10^4$ (top left) the gluon flux eventually catches up
  with, but does not overtake, the $W$ flux. However, extending the
  cascade by another decade by setting $x\subm = 10^5$ (top right),
  gluons do indeed begin to dominate at the smallest values of $x$
  shown.

  The small rate of $A \rightarrow FF$ splittings relative to
  non--Abelian $A \rightarrow AA$ also explains why the spectra of
  $SU(2)$ doublet fermions, which can in principle be produced in the
  first splitting of the original $W$ bosons, do not spike as
  $x \rightarrow x\subm$, in contrast to the fermion spectra in the
  left frame of fig.~\ref{fig:x4BlR}. Here the enhancement of $q_L$
  production compared to splittings to $\ell\subl$, by the color
  factor and the LPM induced relative factor of
  $\alphas/\alpha_\tsub{W}$ is largely canceled by the more efficient
  thermalization of quarks relative to $\ell\subl$, leading to very
  similar normalization of these two spectra at large $x$. At much
  smaller $x$, the flux of $q_L$ (and, for $x\subm \gg 10^4$
  eventually $q_R$) overtakes that of $\ell_L$, due to quark
  production in (tertiary) gluon splitting. Since all $SU(2)$ singlet
  fermions require at least three splitting reactions to be produced,
  e.g. $W \rightarrow \ell_L \rightarrow B \rightarrow \ell_R$, their
  spectra fall of very quickly as $x \rightarrow x\subm$.
 
  The bottom frame of fig.~\ref{fig:x45WlL} shows the non--thermal
  spectra that result from the injection of $SU(2)$ doublet leptons
  $\ell_L$. Here the two possible primary splittings are the emission
  of a $W$ or $B$ boson. Since the corresponding differential rates
  scale like $1/[y^{3/2}\sqrt{1-y}]$ and $1/\sqrt{y(1-y)}$,
    respectively (see eq.(\ref{eq:gamma_qgq})), the emitted gauge
    bosons are predominantly soft, i.e. their spectra do not spike as
    $x \rightarrow x\subm$; clearly soft gauge bosons are even more
    strongly favored in case of $W$ emission, which explains the
    difference between the shapes of the $W$ and $B$ spectra as
    $x \to x\subm$. On the other hand, $SU(2)$ singlet
  (right--chiral) fermions can now already be produced in the second
  step of their cascade, so their spectra at large $x$ lie well above
  those shown in the top frames. In contrast, gluons now require at
  least three splitting reactions to be produced; this, and their very
  short thermalization time, implies that they have by far the
  smallest spectrum as $x \rightarrow x\subm$. This also explains why
  their flux remains below the flux of $W$ bosons even for the
  smallest $x$ shown; however, our earlier arguments show that gluons
  will eventually dominate the cascade if $x\subm \gg 10^4$.

  We finally note that despite similar interactions the shape of the
  $W$ spectrum in the $W$ injection scenario (top frames) differs a
  bit from that of the $\ell_L$ spectrum in the bottom frame. Note
  that the dominant rate for $A \rightarrow AA$ splittings is enhanced
  by a factor $1/(1-y)$ relative to that for $F \rightarrow AF$
  splittings. This favors the emission of relatively harder $W$ bosons
  from an initial $W$, speeding up the loss of very energetic $W$s in
  the cascade. This loss is further enhanced by $A \rightarrow FF$
  splittings which, while rare, reduce the number of gauge bosons in
  the cascade. In contrast, emission off fermions favors very soft
  gauge bosons, leading to smaller energy loss; note also that no
  splitting process can reduce the number of fermions in the
  cascade. These two effects combine to produce a quite pronounced
  minimum in the $W$ spectrum for the case of $W$ injection, at
  $x \sim x\subm/2$, whereas the $\ell_L$ spectrum for the case of
  $\ell_L$ injection reaches its minimum closer to $x\subm$, and rises
  more slowly for smaller $x$. The $x$ dependence of the denominator
  in eq.(\ref{eq:coupleddims}) also plays a role in determining the
  shape of the spectra near their minimum.

\begin{figure}[!ht]
\begin{minipage}{.5\linewidth}
\centering
\includegraphics[scale=.45]{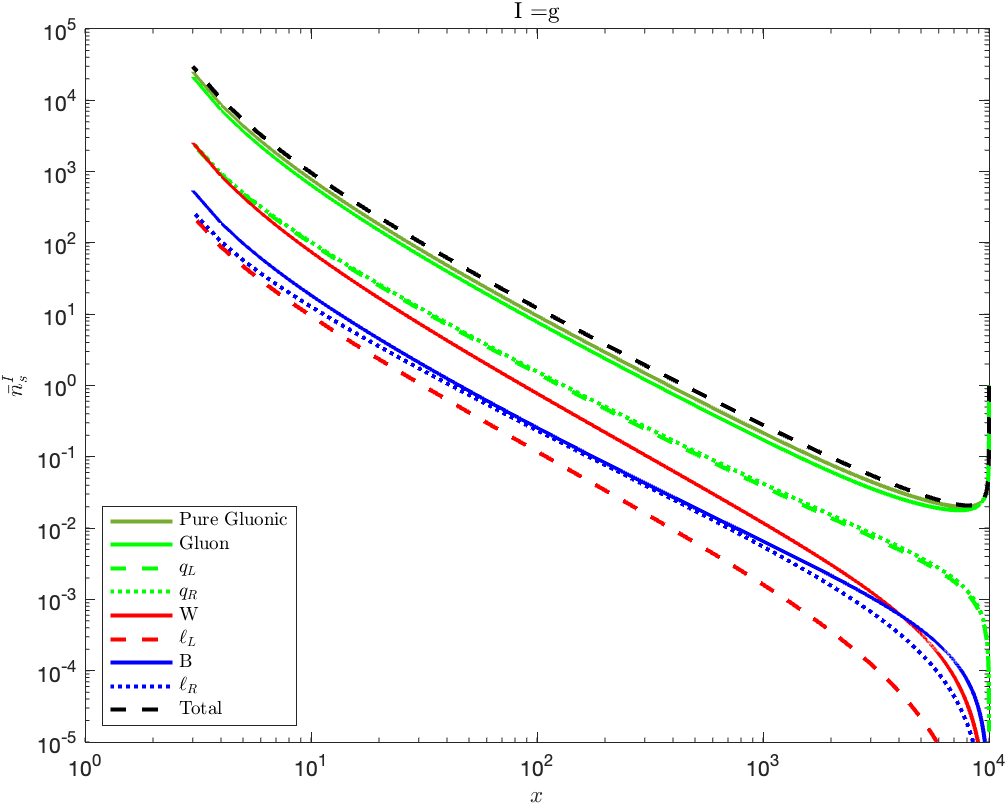}
\end{minipage}
\begin{minipage}{.5\linewidth}
\centering
\includegraphics[scale=.45]{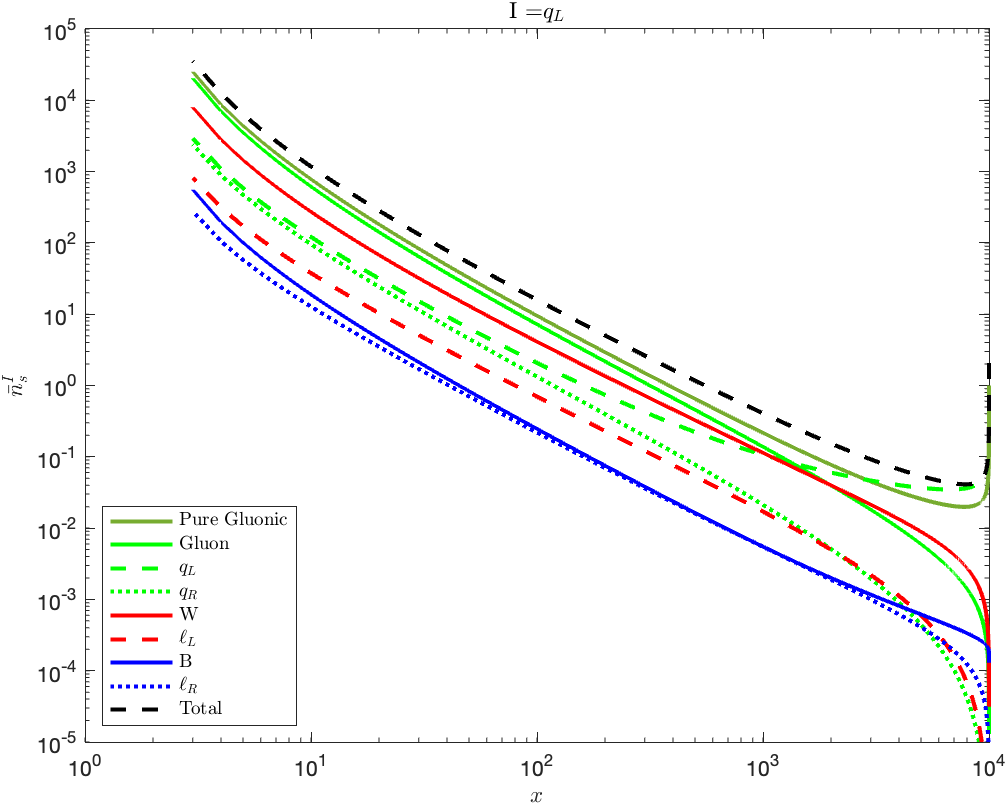}
\end{minipage}\par\medskip
\centering
\includegraphics[scale=.55]{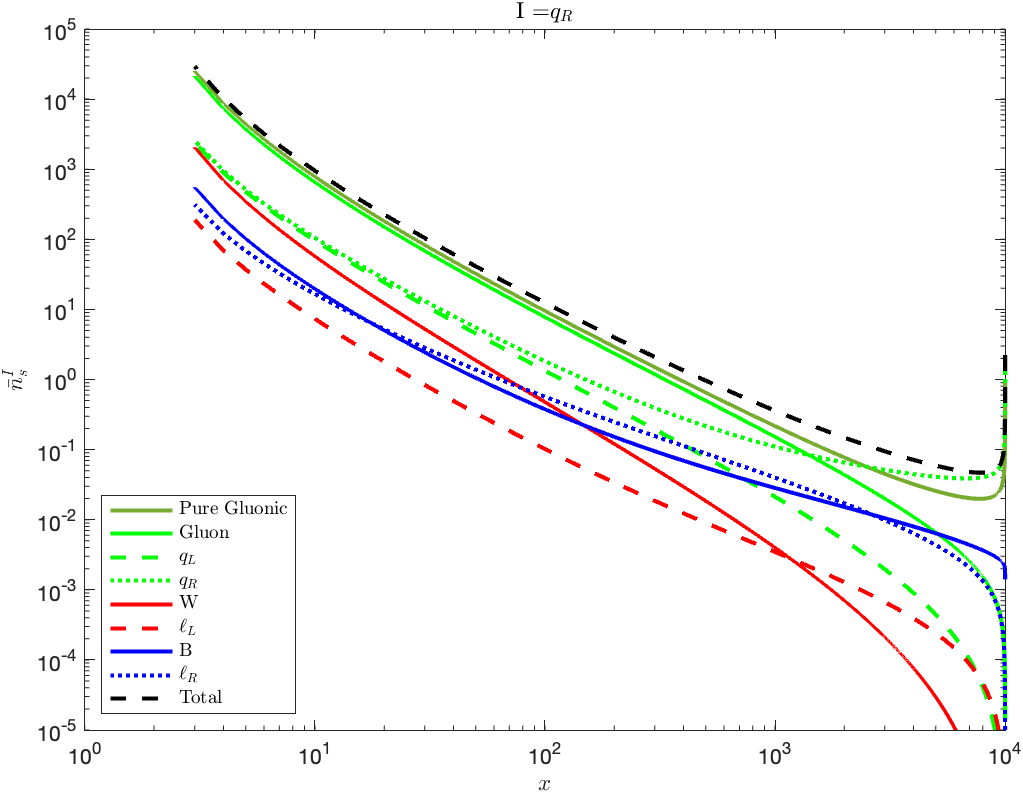}
\caption{The scaled number density functions $\bar{n}_s^I\plr{x}$ of
  eq.(\ref{eq:transfer}) for the various particles $s \in \bbs$, with
  $I=g$ (top Left), $I=q_L$ (top right) and $I=q_R$ (bottom); we
  use $x\subm=10^4$ for all three cases.}
\label{fig:x4gqLqR}
\end{figure}
  
\item \textbf{$I \in \{g,\, q_L,\, q_R\}$}\\
  Finally, let us discuss the cases where the injected particles 
  are charged under $\suthree$. The resulting spectra are shown in
  fig.~\ref{fig:x4gqLqR} for injected gluons (top left), $SU(2)$
  doublet quarks (top right), and $SU(2)$ singlet quarks (bottom).

  Not surprisingly, in all three cases colored particles remain the
  dominant species in the spectrum, with the gluons eventually ending
  up on top. If gluons are the injected particles, their spectrum is
  quite close to that of a pure gluon cascade for all $x$, with a
  slight reduction of the normalization due to the loss of gluons from
  splitting into quarks. This splitting does not distinguish between
  left-- and right--chiral quarks, whose spectra are therefore very
  similar. The latter two almost identical spectra of $q\subl$ and
  $q\subr$ will further source electroweak gauge bosons and leptons,
  appearing first in the second and third step of the cascade,
  respectively. Radiations of $B$ from $q\subr$ will favor harder $B$s
  as discussed above, resulting in a sharper rise of the $B$ spectrum
  at $x \to x\subm$ as compared to that of $W$s. The $W$ spectrum
  overtakes that of the $B$ at lower $x$ despite a larger
  thermalization rate as a result of dominant $W \to WW$ splittings;
  the $B$s on the other hand disappear after each splitting. As
  before, the $\ell\subr$ population is exclusively sourced by, and
  therefore tracks, that of the $B$, with the former winning gradually
  as a result of conserved $\ell\subr$ in splittings. With the $W$
  predominantly sourcing the $W$ and $q\subl$ spectra, only a small
  fraction of all $W$ bosons splits into $\ell\subl$ pairs; this,
  together with having a larger thermalization rate compared to that
  of $\ell\subr$ make the $\ell\subl$ spectrum subdominant.
  
  If quarks are injected, gluons dominate the cascade only for
  $x < x\subm/10$, but then the gluon spectrum again quickly
  approaches that of a pure gluon cascade. Moreover, if $q_L$ are
  injected, $q_R$ production requires at least two splitting
  reactions, and vice versa; as a result, the $q_L$ and $q_R$ spectra
  are very different at large $x$, and converge only for
  $x \leq 10^{-2} x\subm$.

  The chirality of the injected quark is further reflected in the
  large gap between the $W$ and $B$ populations in the two cases: if
  $q_L$ quarks are injected, all electroweak gauge bosons can be
  produced in the first step of the cascade, albeit with significantly
  smaller rates than gluons; in contrast, for $q_R$ injection, $W$
  bosons can only be produced in the third step, e.g.
  $q_R \rightarrow g \rightarrow q_L \rightarrow W$. As a result, in
  the bottom frame $W$ bosons have the smallest flux at large $x$, and
  the flux of $SU(2)$ doublet leptons, which (due to their smaller
  hypercharge) are predominantly produced in the splitting of $W$
  bosons, remain below that of $SU(2)$ singlet leptons over the entire
  range of $x$ shown; note once again that the latter spectrum is also
  enhanced by the smaller thermalization rate of $\ell\subr$.

  Finally, as already discussed in the context of
  fig.~\ref{fig:x45WlL} the shape of the spectrum of the injected
  particle at large $x$ depends to some extent on whether it is a
  fermion of gauge boson species.

\end{itemize}

Figures \ref{fig:x4BlR}, \ref{fig:x45WlL} and \ref{fig:x4gqLqR} show
that for $x$ not much below $x_\tsub{M}$ both the shape and the
normalization of the spectra quite strongly depend on the identity of
the injected particle, i.e. on the branching ratios of the long--lived
matter particles. However, these differences diminish as $x$ becomes
smaller; hence with increasing $x\subm$ a larger and larger part of
the spectrum will be largely independent of the high--scale
$Br_\tsub{I}$ parameters.

\subsection{The role of $x\subm$ and scaling behavior}

In this subsection we quantitatively analyze the role of $x\subm$ in
the normalization and shape of the spectra of non--thermal
particles. We saw in eq.(\ref{eq:nbarsingle}) that for a pure gluon
cascade the normalized spectrum is essentially given by
$1/\sqrt{x\subm}$ times the function \eqref{eq:ffunction} which only
depends on the ratio $x/x\subm$, with minor
corrections.\footnote{Scaling behavior has previously been observed
  both in approximate analytical solutions of our Boltzmann equations
  \cite{Harigaya:2014waa} and in a somewhat different context in
  kinetic theory studies of QGP plasmas (see
  e.g. \cite{Mazeliauskas:2018yef} and references therein); we find it
  reasonable to similarly expect at least an approximate scaling
  behavior from the form of our Boltzmann equations
  \eqref{eq:coupleddims}.} If this holds also for the multi--species
cascade we are analyzing here, the ratios
\begin{equation} \label{eq:ratiofunction}
  r\subs^\tsub{I}\plr{x,x\subm} = \frac { \bar{n}_s^I\plr{x,x\subm} }
  { \bar{n}_{gg}^g\plr{x,x\subm} } 
\end{equation}
should to good approximation only depend on the ratio $x/x\subm$; here
$\bar{n}_{gg}^g$ denotes the pure gluon solution of
\eqref{eq:boltzmannX}, using the full splitting rate of
eq.(\ref{eq:gamma_ggg}).

\begin{figure}[!ht]
\begin{minipage}{.5\linewidth}
\centering
\includegraphics[scale=.45]{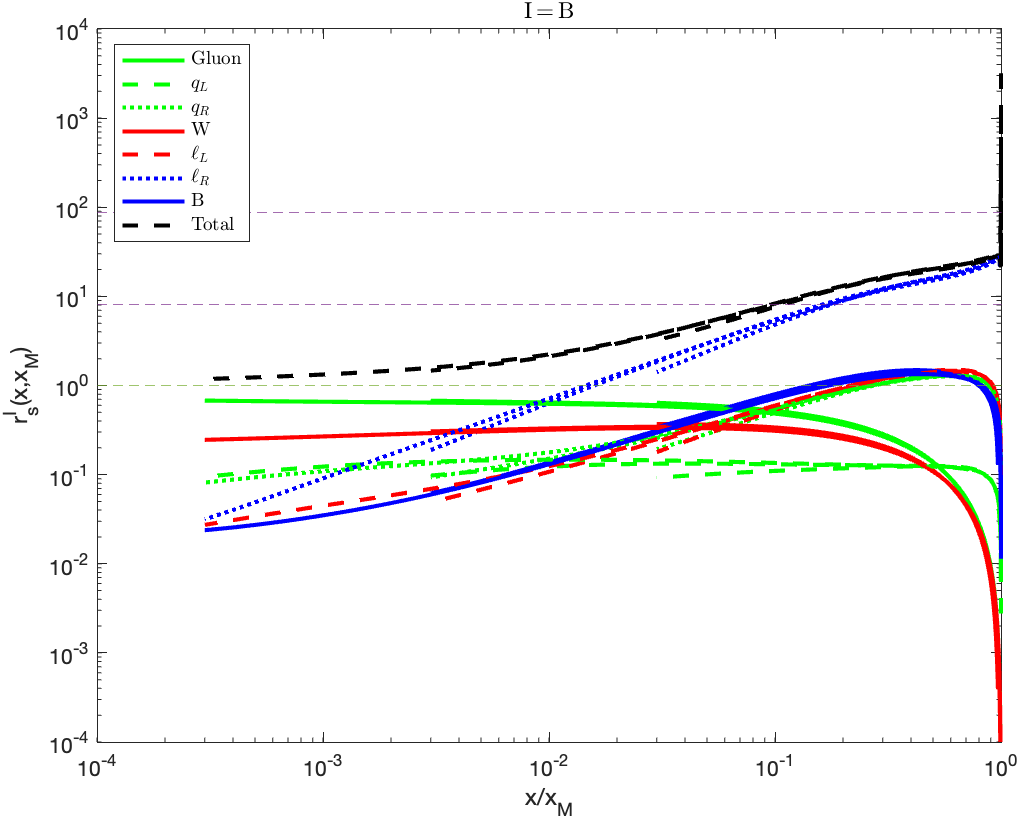}
\end{minipage}
\begin{minipage}{.5\linewidth}
\centering
\includegraphics[scale=.45]{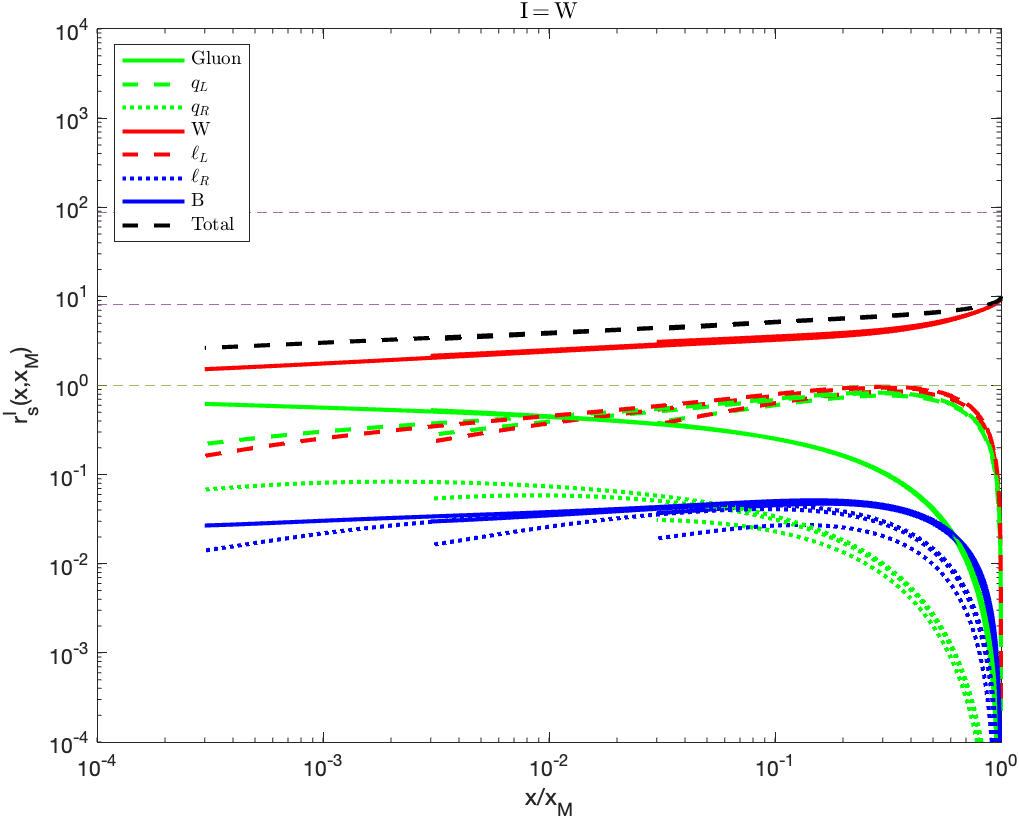}
\end{minipage}\par\medskip
\centering
\includegraphics[scale=.5]{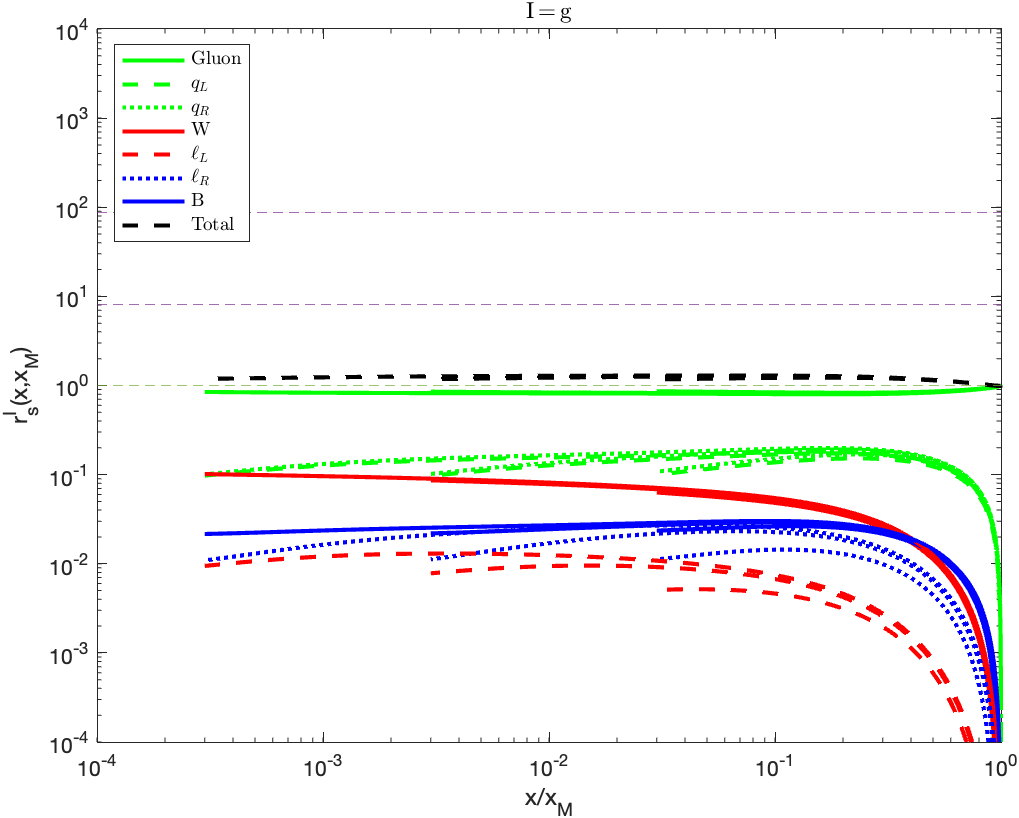}
\caption{Ratios $r_s^I\plr{x,x\subm}$ of eq.(\ref{eq:ratiofunction}),
  for three initial populations: $I=B$ (top left), $W$ (top right),
  and $g$ (bottom). Results for $x\subm=10^2,\, 10^3,\, 10^4$ are
  overlaid. The dark green horizontal dashed line marks $r=1$,
  corresponding to a pure gluon cascade. The two purple dashed lines
  show the ratios
  $\frac{\Gamma^{\rm split}_\tsub{LPM}\plr{g,x\subm}}{\Gamma^{\rm
      split}_\tsub{LPM}\plr{W,x\subm}}$, and
  $\frac{\Gamma^{\rm split}_\tsub{LPM}\plr{g,x\subm}}{\Gamma^{\rm
      split}_\tsub{LPM}\plr{B,x\subm}}$, see
  eq.(\ref{eq:Sec2-dimlesstotalrate}), which determine the
  normalizations of the spectra of injected $W$ and $B$ bosons,
  respectively, for $x \rightarrow x\subm$.}
\label{fig:transferbwg}
\end{figure}

Some results are shown in fig.~\ref{fig:transferbwg}. Here we focus on
scenarios where the injected particles are gauge bosons, $I=g,\, W,\,$
and $B$, and overlaid curves for three values of
$x\subm=10^2,\, 10^3,\,$ and $10^4$. We see that the rescaled spectra
of gauge bosons indeed to very good approximation only depend on
$x/x\subm$. We saw in the discussion of fig.~\ref{fig:x4BlR} that the
evolution equations are quite similar for gluons and for $W$ bosons;
it is therefore not surprising that the rescaled $W$ spectra $r_W$
don't show a strong dependence on $x\subm$, and become quite flat at
$x \ll x\subm$. The evolution equation for Abelian $B$ bosons is
quite different, however, due to the absence of $B \rightarrow BB$
splitting and the strong LPM suppression of $F \rightarrow B F$
splitting. It is therefore not surprising that $r_B$ can have quite a
different shape than $r_W$; the fact that $r_B$ to very good approximation
depends on $x\subm$ only via the ratio $x/x\subm$ appears to be
\enquote{accidental}.

In our results, the rescaled fermion spectra do show some residual dependence on $x\subm$.
This is especially pronounced for $r_{\ell_R}$; right--chiral leptons
can lose energy only via emission of $B$ bosons. Recall that the
corresponding differential splitting rate only scales like $1/\sqrt{y}$
for small $y$, while all rates for emitting a non--Abelian gauge boson
scale like $1/y^{3/2}$ in that limit. $\ell_R$ therefore typically
emits more energetic gauge bosons, leading to a significantly shorter
cascade\footnote{More exactly, a cascade containing fewer splittings, which
  nevertheless takes more time, due to the small total splitting rate
  of $l_R$.} and hence a stronger dependence on the boundary conditions.

The other rescaled fermion spectra explicitly depend on $x\subm$
mostly at relatively small $x$. This may reflect the effect of
$A \rightarrow FF$ splitting on the total splitting rate of $A$ which,
as we saw above, becomes relatively more important at smaller $x$.
The same effect also explains why the explicit $x\subm$ dependence is
somewhat stronger for left--chiral leptons than for quarks; we saw in
the discussion of fig.~\ref{fig:x4BlR} that $A \rightarrow FF$
splitting is more important for $A = W$ than for $A=g$.

Figure \ref{fig:transferbwg} also illustrates the flow towards a
non--thermal spectrum dominated by gluons and quarks, irrespective of
the initial injection. In particular, in the case of gluon injection,
the ratio functions settle quickly into their asymptotic values, so
that the particle ratios depend only weakly on $x$ for
$x \leq 0.1 x\subm$. At least for this scenario, and for the given values
of the gauge couplings, one should therefore
be able to predict the various ratio functions for $x\subm > 10^4$
quite accurately by extrapolating the numerical results of the bottom
frame of fig.~\ref{fig:transferbwg}, without the need for new numerical
solutions of eqs.(\ref{eq:coupleddims}); recall that the gauge couplings do depend weakly on
$T$, and hence on $x\subm$ for fixed $M$. Not surprisingly, if the
injected particle is a color singlet, a longer cascade is required
before strongly interacting particles dominate.

\begin{figure}[ht!]
\centering
\includegraphics[width=0.6 \textwidth]{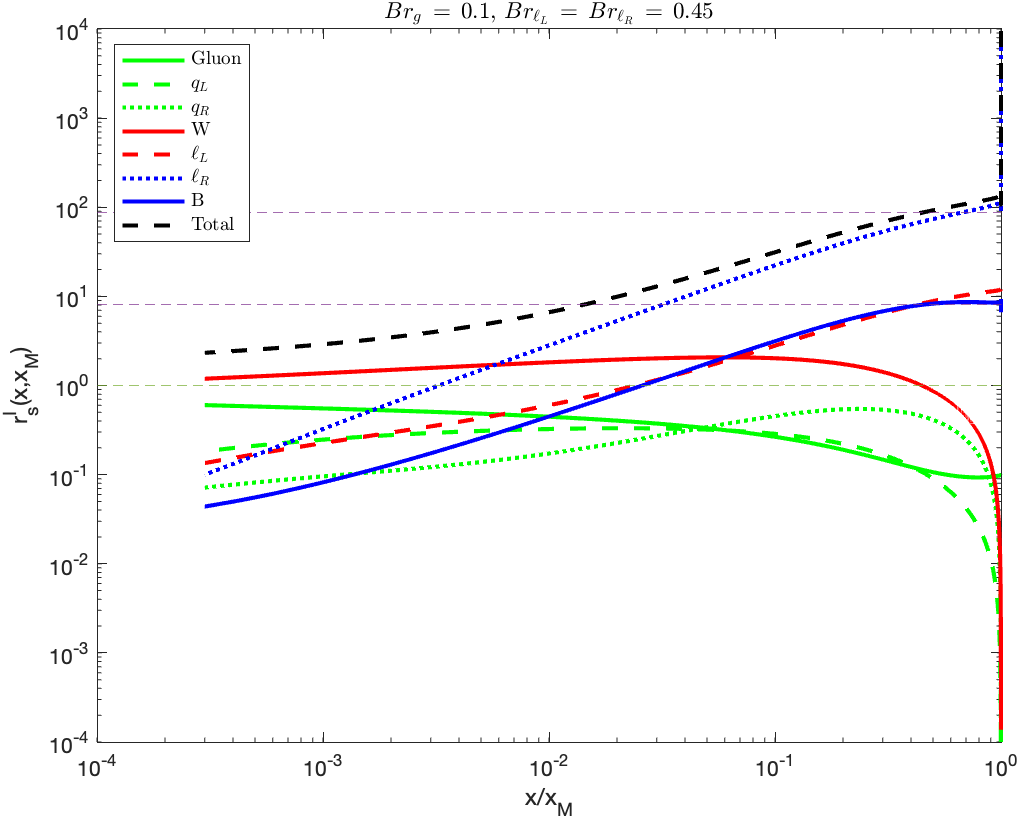}
\caption{Ratios $r_s\plr{x,x\subm}$ for a scenario where several
  species are injected, with branching ratios
  $Br_g = 0.1,\, Br_{\ell_L} = Br_{\ell_R}=0.45$, with
  $x\subm=10^4$. The dashed horizontal lines correspond to an initial
  $W$ and $B$ injection as explained in fig.~\ref{fig:transferbwg}.}
\label{fig:transfercheck}
\end{figure}

Finally, once all $\bar{n}_s^I\plr{x}$ have been computed one can use
eq.(\ref{eq:transfer}) to deduce the spectrum of non--thermal
particles for any given set of branching ratios $Br_I$. One such
example is shown in fig.~\ref{fig:transfercheck}, where we assumed
$Br_g = 0.1,\, Br_{\ell_L} = Br_{\ell_R} = 0.45$ with $x\subm=10^4$;
we checked explicitly that a direct numerical solution of
eqs.(\ref{eq:coupleddims}) gives the same result. We see that the
spectrum remains dominated by $\ell_R$ for most of the range of $x$;
this reflects the very large normalization of the $\ell_R$ spectrum,
as discussed previously for fig.~\ref{fig:x4BlR}.

\section{Application: semi-thermal production of heavy dark matter}
\label{sec:application}

As an example of the application of our results, we briefly discuss
the production of heavy particles $\chi$ via annihilation reactions
involving one non--thermal particle and one particle from the thermal
bath. For definiteness, we will assume that the long--lived particles,
with mass $M$ and decay width $\Gamma\subm$, whose decays were
ultimately responsible for the spectra of non--thermal particles,
dominates the energy density of the universe prior to its decay. Moreover, we will focus
on the case where the lifetime of the produced particles is much
larger than $1/\Gamma_{\rm M}$, so that we can treat them as stable;
this is certainly true if we are interested in the production of dark
matter.

The long--lived very massive particles dominated the energy density as
long as the temperature is above the reheating temperature, which is
given by
\begin{equation} \label{eq:reheattemp}
T_\tsub{RH} \sim  \sqrt{\Gamma\subm M_\tsub{Pl}}\,,
\end{equation}
where $M_\tsub{Pl}$ is the reduced Planck mass. If
$m_\chi \lsim T_\tsub{RH}$, total $\chi$ pair production through
renormalizable interactions will be dominated by reactions only
involving thermalized particles, so the resulting $\chi$ density will
be insensitive to the details of the spectrum of nonthermal
particles. On the other hand, if $m_{\chi} \gg 20 T_\tsub{RH}$,
thermal production of $\chi$ particles with roughly weak--strength
cross sections (e.g. the well--known WIMPs \cite{Bertone:2004pz}) will
effectively cease at temperature
$T_\tsub{P} \sim \order{0.05 m_{\chi}} > T_\tsub{RH}$; a possible
thermal contribution to the $\chi$ density will then be diluted by the
entropy production between $T_\tsub{P}$ and $T_\tsub{RH}$. In this
case the final $\chi$ density may well be dominated by non--thermal
(or semi--thermal) processes of the kind
\begin{equation} \label{eq:xprod}
s\, s' \longrightarrow \chi \, \chi\,,
\end{equation}
where $s'$ denotes a particle from the thermal bath but $s$ is taken
from the spectrum of non--thermal particles, with momentum
$p_s \gg T$. We denote the corresponding production cross section by
$\sigma_{\chi}^\tsub{ST}$. The $\chi$ number density $n_\chi$ then
obeys the equation
\begin{eqnarray} \label{eq:semithermprodnumden}
  \frac{d n_{\chi}^\tsub{ST} \plr{T} } {dt} + 3 H n_\chi^\tsub{ST}
  &=& \sum_{s, \, s' }
\int n_{s'}\plr{T} \langle \sigma_{\chi}^\tsub{ST}v \rangle \plr{p,T} \tilde n_s
   \plr{p,T} dp \nonumber \\ 
 &=& \sum_{s, \, s' } \int n_{s'}\plr{T} \langle \sigma_{\chi}^\tsub{ST}
v \rangle  \plr{p,T} r_s \plr{p,T} \tilde n_{gg}^g \plr{p,T} dp\,.
\end{eqnarray}
Here $n_{s'}\plr{T} = c_{s'} \zeta(3) g_{s'} T^3/\pi^2$ with
$c_{s'} = 1 \ (3/4)$ for bosonic (fermionic) $s'$ is the thermal
number density of the species $s'$ with $g_{s'}$ degrees of freedom,
$\langle \sigma_{\chi}^\tsub{ST}v \rangle$ denotes the thermal average
of the production cross section, and we have used
eq.(\ref{eq:multinormalization}) to write $\tilde n_s\plr{p,T}$ in
terms of the pure gluon solution of eq.(\ref{eq:boltzmannX}). In
eq.(\ref{eq:semithermprodnumden}) we have assumed that $n_\chi$
remains so small that $\chi$ annihilation processes are irrelevant.
While analytic approximations and numerical solutions for the pure
gauge solution have previously been used to estimate the $\chi$ number
density, our knowledge of the ratios $r\subs$ allows for a much more
precise calculation for the production rate of the process
\eqref{eq:xprod}.

Of course, $\chi$ pair production is only possible if the center--of--mass 
energy exceeds $2 m_\chi$. Since the average energy of a particle
in the thermal bath is around $3T$ and the spectrum of non--thermal
particles quickly increases with decreasing momentum, the dominant
contribution to the integral in eq.(\ref{eq:semithermprodnumden})
typically comes from $p \sim m_\chi^2/T$. Together with the fact that
$n_{s'} \propto T^3$ in eq.(\ref{eq:semithermprodnumden}), this
implies that the $\chi$ production rate quickly increases with
increasing $T$. However, the entropy dilution factor
\cite{Drees:2021lbm} $(T_{\rm RH}/T)^5$ implies that nevertheless the
biggest contribution to the final $\chi$ density often comes from
temperature $T \simeq T_{\rm RH}$, unless $m_\chi^2 > T_{\rm RH} M$.
If $m_\chi^2 \ll T_{\rm RH} M$ the semi--thermal production rate will
be dominated by particles with energy $p \ll M$, whose spectrum is
less dependent on the decay branching ratios of our heavy matter
particles. In contrast, for $m_\chi^2 \gsim T_{\rm RH} M$, very
energetic particles with $p \sim M/2$ will make sizable or even
dominant contributions; their spectra do strongly depend on the
initial branching ratios. As long as $m_\chi^2 < T_{\rm max} M$,
where $T_{\rm max}$ is the maximal temperature of the early matter
dominated epoch, most $\chi$ particles will be produced at temperature
$T \sim m_\chi^2/M$.

Finally, the production of very massive particles, with
$m_\chi^2 \gsim T_{\rm max} M$, will be dominated by purely
non--thermal processes, where {\em both} initial particles in the
reaction \eqref{eq:xprod} are non--thermal \cite{Allahverdi:2002pu,
  Harigaya:2019tzu, Drees:2021lbm}. The distinction between the
various species is in that case of even greater importance as now the
product of ratios $r_s r_{s'}$ appears. We leave a more careful study
of dark matter production from different phases of a generic matter
domination era for a future work.

\section{Summary and discussion}
\label{sec:conclusion}

Cosmological histories including an early matter domination epoch,
where the energy density of the universe is dominated by a long--lived
matter component of mass $M$ and decay width $\Gamma\subm$, are well
motivated and have been studied in many recent publications. The
inclusion of such a period has been shown to potentially affect
various aspects of cosmology, including the production of dark matter
or the baryon asymmetry. Since we know that the universe was dominated
by the SM radiation bath at the latest at the onset of BBN, any matter
component should decay and eventually (predominantly) thermalize into
the thermal bath.

The thermalization procedure entails the splitting of initial decay
products of energy $p_i \gg T$ into $\order{p_i/T}$ particles of
energy $\order{T}$, and has been studied in the literature both in
cosmological contexts and in the context of heavy ion collisions. So
long as the out of equilibrium particles are gauge charged, the
dominant process of thermalization is known to be the $1 \to 2$ near
collinear splittings of the energetic particles made possible via soft
$t-$channel interactions with the thermal bath, mediated by gauge
bosons with thermal mass $\order{gT}$, $g$ being the relevant gauge
coupling. The collinearity of the splitting then calls for a careful
inclusion of the LPM effect resulting from consecutive interactions of
the particles with the plasma during the formation time of the
splitting process. The LPM suppressed rate of pure--gauge splittings
$g \to gg$ had already been studied analytically and numerically in
the literature; these studies are motivated by the fact that the QCD
sector, and in particular gluons, can be expected to dominate the
spectrum of nonthermal particles, at least for momenta $p \ll p_i$.

In this study, we extend the previous works by including the full set
of chiral SM fermions and gauge bosons in the splitting cascade of an
initial population of energetic particles of energy $p_i=M/2$. The
functions describing the emission of a gauge boson all have similar
functional form, but can have different coupling strengths, flavor
multiplicity factors, and numerical LPM suppression factors; evidently
these processes increase the number of gauge bosons. In contrast, the
splitting of a gauge boson into an $f \bar f$ pair is suppressed by a
factor $p_d/p_p$, where $p_d$ and $p_p$ denote the momentum of the
softer daughter and parent particle, respectively; it increases the
number of fermions, but reduces the number of gauge bosons. Note also
that this is the only splitting process available to Abelian gauge
bosons.

We use the explicitly calculated results for LPM suppressed splitting
rates corresponding to an $SU(N)$ gauge group from the literature, and
use physical arguments to deduce the rate for processes involving
species charged under different gauge groups, see
eqs.(\ref{eq:dGamma}) to (\ref{eq:uonegroupfacs}). The resulting
Boltzmann equations can be written in terms of the dimensionless ratio
$x = p/T$, see eq.(\ref{eq:coupleddims}); in this form the equations
depend only on $x\subm = M/2T$ rather than on $M$ and $T$
separately. Of course, the decay branching ratios of the matter particles
into the various SM species are also important, but due to the
linearity of the problem, we only need to consider the limiting cases
where decay into a single species dominates. The corresponding
numerical results are shown in figures~\ref{fig:x4BlR} to
\ref{fig:x4gqLqR}.

We find that spectra of individual species can have order of magnitude
deviations from that of the pure--gluon solution over several decades
of momentum, i.e. for $1 \lsim x/x\subm \lsim 10^3$. We treat the
fermion chiralities separately, since only left--chiral fermions (and
right--chiral antifermions) have $SU(2)$ interaction. In scenarios
with an initial chiral asymmetry in the matter decay products, it also
persists for several decades in $x$. On the other hand, an approximate
scale invariant behavior is observed for the ratios of the various
species \eqref{eq:ratiofunction}, which for $x \ll x\subm$
asymptotically approach unique solutions independent of the branching
ratios.

Our treatment should suffice for many practical purposes, greatly
improving the precision of the calculation of cosmological processes
involving non--thermal particles. The validity of the results is
however limited by the approximations we made. For example, in writing
the Boltzmann equations \eqref{eq:coupleddims}, we disregard the role
of hard $2 \to 2$ processes. This should be a good approximation so
long as $x \gg 1$\footnote{Recently, interpolation schemes have been
  suggested \cite{Ghiglieri:2021vcq} to smoothly cross over to the
  unsuppressed Bethe--Heitler regime of eq.(\ref{eq:bethe}).}, but the
low$-x$ tail of our result could be affected. On the other hand, these
results for the ratios of spectra, together with the pure gluon
solution, should serve as a good approximation for $x \gg 1$ regions
even for cases with $x\subm \gg 10^4$ as we argued in section
\ref{sec:solution}. Recall also that the total flux of particles with
momentum $p \lsim 10 T$ will in any case be dominated by the thermal
component.

Moreover, we did not allow for the production of scalar particles in
the cascade. In section~\ref{subsec:particles} we argued that we may
safely leave out the SM Higgs doublet in our analysis, but this may be
different in models with extended Higgs sectors (with more degrees of
freedom, and often also enhanced couplings to some fermions). Scalars,
and Majorana gauginos, will certainly be important if one wishes to
analyze a supersymmetric cascade. Results for medium induced
scalar--scalar--gauge boson splittings have been published in the
literature \cite{Arnold:2002ja}; these would need to be augmented with
splitting functions involving Yukawa interactions.


\begin{thebibliography}{99}
%

\bibitem{kt}
E.~W.~Kolb and M.~S.~Turner, ``The Early Universe,''
Front. Phys.\textbf{69} (1990).

\bibitem{Allahverdi:2010xz}
R.~Allahverdi, R.~Brandenberger, F.~Y.~Cyr-Racine and A.~Mazumdar,
``Reheating in Inflationary Cosmology: Theory and Applications,''
Ann. Rev. Nucl. Part. Sci. \textbf{60} (2010), 27, 
doi:10.1146/annurev.nucl.012809.104511
[arXiv:1001.2600 [hep-th]].

\bibitem{Kane:2015jia}
G.~Kane, K.~Sinha and S.~Watson,
``Cosmological Moduli and the Post-Inflationary Universe: A Critical Review,''
Int. J. Mod. Phys. D \textbf{24} (2015) no.08, 1530022,
doi:10.1142/S0218271815300220
[arXiv:1502.07746 [hep-th]].

\bibitem{Allahverdi:2020bys}
R.~Allahverdi et al., 
``The First Three Seconds: a Review of Possible Expansion Histories of the
Early Universe,'' 
doi:10.21105/astro.2006.16182
[arXiv:2006.16182 [astro-ph.CO]].

\bibitem{Giudice:2000ex}
G.~F.~Giudice, E.~W.~Kolb and A.~Riotto,
``Largest temperature of the radiation era and its cosmological implications,''
Phys. Rev. D \textbf{64} (2001), 023508
doi:10.1103/PhysRevD.64.023508
[arXiv:hep-ph/0005123 [hep-ph]].

\bibitem{Hannestad:2004px}
S.~Hannestad,
``What is the lowest possible reheating temperature?'',
Phys. Rev. D \textbf{70} (2004), 043506
doi:10.1103/PhysRevD.70.043506
[arXiv:astro-ph/0403291 [astro-ph]].

\bibitem{KaneMaybeMatter}
  J.~T.~Giblin, G.~Kane, E.~Nesbit, S.~Watson and Y.~Zhao,
  ``Was the Universe Actually Radiation Dominated Prior to Nucleosynthesis?,''
  Phys.\ Rev.\ D {\bf 96} (2017) no.4,  043525
  doi:10.1103/PhysRevD.96.043525
  [arXiv:1706.08536 [hep-th]].

\bibitem{Berlin:2016vnh}
A.~Berlin, D.~Hooper and G.~Krnjaic,
``PeV-Scale Dark Matter as a Thermal Relic of a Decoupled Sector,''
Phys. Lett. B \textbf{760} (2016), 106-111
doi:10.1016/j.physletb.2016.06.037
[arXiv:1602.08490 [hep-ph]].

\bibitem{Berlin:2016gtr}
A.~Berlin, D.~Hooper and G.~Krnjaic,
``Thermal Dark Matter From A Highly Decoupled Sector,''
Phys. Rev. D \textbf{94} (2016) no.9, 095019
doi:10.1103/PhysRevD.94.095019
[arXiv:1609.02555 [hep-ph]].
  
\bibitem{Chung:1998rq}
D.~J.~H.~Chung, E.~W.~Kolb and A.~Riotto,
``Production of massive particles during reheating,''
Phys. Rev. D \textbf{60} (1999), 063504,
doi:10.1103/PhysRevD.60.063504
[arXiv:hep-ph/9809453 [hep-ph]].

\bibitem{AD2}
R.~Allahverdi and M.~Drees,
``Production of massive stable particles in inflaton decay,''
Phys. Rev. Lett. \textbf{89} (2002), 091302, 
doi:10.1103/PhysRevLett.89.091302
[arXiv:hep-ph/0203118 [hep-ph]].
  
\bibitem{Allahverdi:2002pu}
  R.~Allahverdi and M.~Drees,
  ``Thermalization after inflation and production of massive stable particles,''
  Phys.\ Rev.\ D {\bf 66} (2002) 063513
  doi:10.1103/PhysRevD.66.063513
  [hep-ph/0205246].
  
\bibitem{Gelmini:2006pw}
  G.~B.~Gelmini and P.~Gondolo,
  ``Neutralino with the right cold dark matter abundance in (almost) any
  supersymmetric model'',
  Phys.\ Rev.\ D {\bf 74} (2006) 023510
  doi:10.1103/PhysRevD.74.023510
  [hep-ph/0602230].

\bibitem{Acharya:2008bk}
B.~S.~Acharya, P.~Kumar, K.~Bobkov, G.~Kane, J.~Shao and S.~Watson,
``Non-thermal Dark Matter and the Moduli Problem in String Frameworks'',
JHEP \textbf{06} (2008), 064, doi:10.1088/1126-6708/2008/06/064
[arXiv:0804.0863 [hep-ph]].

\bibitem{Kane:2011ih}
G.~Kane, J.~Shao, S.~Watson and H.~B.~Yu,
``The Baryon-Dark Matter Ratio Via Moduli Decay After Affleck-Dine
Baryogenesis'', 
JCAP \textbf{11} (2011), 012, doi:10.1088/1475-7516/2011/11/012
[arXiv:1108.5178 [hep-ph]].

\bibitem{Hasenkamp:2012ii}
J.~Hasenkamp and J.~Kersten,
``Dark radiation from particle decay: cosmological constraints and
opportunities'',
JCAP \textbf{08} (2013), 024, doi:10.1088/1475-7516/2013/08/024
[arXiv:1212.4160 [hep-ph]].

\bibitem{Kurata:2012nf}
  Y.~Kurata and N.~Maekawa,
  ``Averaged Number of the Lightest Supersymmetric Particles in Decay of
  Superheavy Particle with Long Lifetime'',
  Prog.\ Theor.\ Phys.\  {\bf 127} (2012) 657,
  doi:10.1143/PTP.127.657
  [arXiv:1201.3696 [hep-ph]].

\bibitem{Harigaya:2014waa}
  K.~Harigaya, M.~Kawasaki, K.~Mukaida and M.~Yamada,
  ``Dark Matter Production in Late Time Reheating'',
  Phys.\ Rev.\ D {\bf 89} (2014) no.8, 083532,
  doi:10.1103/PhysRevD.89.083532
  [arXiv:1402.2846 [hep-ph]].

\bibitem{Ishiwata:2014cra}
K.~Ishiwata,
``Axino Dark Matter in Moduli-induced Baryogenesis'',
JHEP \textbf{09} (2014), 122, doi:10.1007/JHEP09(2014)122
[arXiv:1407.1827 [hep-ph]].

\bibitem{Kane:2015qea}
G.~L.~Kane, P.~Kumar, B.~D.~Nelson and B.~Zheng,
``Dark matter production mechanisms with a nonthermal cosmological history:
A classification'', 
Phys. Rev. D \textbf{93} (2016) no.6, 063527
doi:10.1103/PhysRevD.93.063527 [arXiv:1502.05406 [hep-ph]].

\bibitem{Co:2015pka}
  R.~T.~Co, F.~D'Eramo, L.~J.~Hall and D.~Pappadopulo,
  ``Freeze-In Dark Matter with Displaced Signatures at Colliders'',
  JCAP {\bf 1512} (2015) no.12, 024, doi:10.1088/1475-7516/2015/12/024
  [arXiv:1506.07532 [hep-ph]].

\bibitem{Dhuria:2015xua}
M.~Dhuria, C.~Hati and U.~Sarkar,
``Moduli induced cogenesis of baryon asymmetry and dark matter'',
Phys. Lett. B \textbf{756} (2016), 376-383
doi:10.1016/j.physletb.2016.03.018
[arXiv:1508.04144 [hep-ph]].

\bibitem{Hamdan:2017psw}
S.~Hamdan and J.~Unwin,
``Dark Matter Freeze-out During Matter Domination'',
Mod. Phys. Lett. A \textbf{33} (2018) no.29, 1850181,
doi:10.1142/S021773231850181X
[arXiv:1710.03758 [hep-ph]].

\bibitem{Kim:2016spf}
  H.~Kim, J.~P.~Hong and C.~S.~Shin,
  ``A map of the non-thermal WIMP,''
  Phys.\ Lett.\ B {\bf 768} (2017) 292
  doi:10.1016/j.physletb.2017.03.005
  [arXiv:1611.02287 [hep-ph]].

\bibitem{Drees:2017iod}
  M.~Drees and F.~Hajkarim,
  ``Dark Matter Production in an Early Matter Dominated Era'',
  JCAP {\bf 1802} (2018) no.02, 057, doi:10.1088/1475-7516/2018/02/057
  [arXiv:1711.05007 [hep-ph]].

\bibitem{Garcia:2018wtq}
M.~A.~G.~Garcia and M.~A.~Amin,
``Prethermalization production of dark matter'',
Phys. Rev. D \textbf{98} (2018) no.10, 103504,
doi:10.1103/PhysRevD.98.103504 [arXiv:1806.01865 [hep-ph]].

\bibitem{Drees:2018dsj}
  M.~Drees and F.~Hajkarim,
  ``Neutralino Dark Matter in Scenarios with Early Matter Domination,''
  JHEP {\bf 1812} (2018) 042
  doi:10.1007/JHEP12(2018)042
  [arXiv:1808.05706 [hep-ph]].

\bibitem{Chanda:2019xyl}
P.~Chanda, S.~Hamdan and J.~Unwin,
``Reviving $Z$ and Higgs Mediated Dark Matter Models in Matter Dominated
Freeze-out'',
JCAP \textbf{01} (2020), 034, doi:10.1088/1475-7516/2020/01/034
[arXiv:1911.02616 [hep-ph]].
  
\bibitem{Maldonado:2019qmp}
C.~Maldonado and J.~Unwin,
``Establishing the Dark Matter Relic Density in an Era of Particle Decays,''
JCAP \textbf{06} (2019), 037
doi:10.1088/1475-7516/2019/06/037
[arXiv:1902.10746 [hep-ph]].

\bibitem{Harigaya:2019tzu}
  K.~Harigaya, K.~Mukaida and M.~Yamada,
  ``Dark Matter Production during the Thermalization Era,''
  arXiv:1901.11027 [hep-ph].

\bibitem{Garcia:2020eof}
M.~A.~G.~Garcia, K.~Kaneta, Y.~Mambrini and K.~A.~Olive,
Phys. Rev. D \textbf{101} (2020) no.12, 123507
doi:10.1103/PhysRevD.101.123507
[arXiv:2004.08404 [hep-ph]].
  
\bibitem{Hamada:2015xva}
Y.~Hamada and K.~Kawana,
``Reheating-era leptogenesis,''
Phys. Lett. B \textbf{763} (2016), 388-392
doi:10.1016/j.physletb.2016.10.067
[arXiv:1510.05186 [hep-ph]].

\bibitem{Asaka:2019ocw}
T.~Asaka, H.~Ishida and W.~Yin,
``Direct baryogenesis in the broken phase,''
JHEP \textbf{07} (2020), 174, doi:10.1007/JHEP07(2020)174
[arXiv:1912.08797 [hep-ph]].

 \bibitem{Davidson:2000er}
S.~Davidson and S.~Sarkar,
``Thermalization after inflation,''
JHEP \textbf{11} (2000), 012, doi:10.1088/1126-6708/2000/11/012
[arXiv:hep-ph/0009078 [hep-ph]].

\bibitem{Harigaya:2013vwa}
 K.~Harigaya and K.~Mukaida, ``Thermalization after/during Reheating,''
JHEP {\bf 1405} (2014) 006, doi:10.1007/JHEP05(2014)006
  [arXiv:1312.3097 [hep-ph]].

\bibitem{Drees:2021lbm}
M.~Drees and B.~Najjari,
``Energy Spectrum of Thermalizing High Energy Decay Products in the Early Universe,''
JCAP \textbf{10} (2021), 009,
doi:10.1088/1475-7516/2021/10/009 [arXiv:2105.01935 [hep-ph]].

\bibitem{Landau:1953um}
L.~D.~Landau and I.~Pomeranchuk,
``Limits of applicability of the theory of bremsstrahlung electrons and pair production at high-energies,''
Dokl.\ Akad.\ Nauk Ser.\ Fiz.\  {\bf 92} (1953) 535.

\bibitem{Migdal:1956tc}
 A.~B.~Migdal,
``Bremsstrahlung and pair production in condensed media at high-energies,''
Phys.\ Rev.\  {\bf 103} (1956) 1811,  doi:10.1103/PhysRev.103.1811

\bibitem{Baier:1998kq}
R.~Baier, Y.~L.~Dokshitzer, A.~H.~Mueller and D.~Schiff,
``Medium induced radiative energy loss: Equivalence between the BDMPS and Zakharov formalisms,''
Nucl. Phys. B \textbf{531} (1998), 403, doi:10.1016/S0550-3213(98)00546-X
[arXiv:hep-ph/9804212 [hep-ph]].

\bibitem{Arnold:2002zm}
P.~B.~Arnold, G.~D.~Moore and L.~G.~Yaffe,
``Effective kinetic theory for high temperature gauge theories,''
JHEP \textbf{01} (2003), 030, doi:10.1088/1126-6708/2003/01/030
[arXiv:hep-ph/0209353 [hep-ph]].

\bibitem{Jeon:2003gi}
S.~Jeon and G.~D.~Moore,
``Energy loss of leading partons in a thermal QCD medium,''
Phys. Rev. C \textbf{71} (2005), 034901, doi:10.1103/PhysRevC.71.034901
[arXiv:hep-ph/0309332 [hep-ph]].

\bibitem{Arnold:2008zu}
P.~B.~Arnold and C.~Dogan,
``QCD Splitting/Joining Functions at Finite Temperature in the Deep LPM
Regime,''
Phys. Rev. D \textbf{78} (2008), 065008
doi:10.1103/PhysRevD.78.065008 [arXiv:0804.3359 [hep-ph]].

\bibitem{Kurkela:2011ti}
A.~Kurkela and G.~D.~Moore,
``Thermalization in Weakly Coupled Nonabelian Plasmas,''
JHEP \textbf{12} (2011), 044, doi:10.1007/JHEP12(2011)044
[arXiv:1107.5050 [hep-ph]].

\bibitem{AbraaoYork:2014hbk}
M.~C.~Abraao York, A.~Kurkela, E.~Lu and G.~D.~Moore,
``UV cascade in classical Yang-Mills theory via kinetic theory,''
Phys. Rev. D \textbf{89} (2014) no.7, 074036
doi:10.1103/PhysRevD.89.074036
[arXiv:1401.3751 [hep-ph]].

\bibitem{Arnold:2001ba}
P.~B.~Arnold, G.~D.~Moore and L.~G.~Yaffe,
``Photon emission from ultrarelativistic plasmas,''
JHEP \textbf{11} (2001), 057, doi:10.1088/1126-6708/2001/11/057
[arXiv:hep-ph/0109064 [hep-ph]].

\bibitem{Arnold:2001ms}
P.~B.~Arnold, G.~D.~Moore and L.~G.~Yaffe,
``Photon emission from quark gluon plasma: Complete leading order results,''
JHEP \textbf{12} (2001), 009, doi:10.1088/1126-6708/2001/12/009
[arXiv:hep-ph/0111107 [hep-ph]].

\bibitem{Arnold:2002ja}
P.~B.~Arnold, G.~D.~Moore and L.~G.~Yaffe,
``Photon and gluon emission in relativistic plasmas,''
JHEP \textbf{06} (2002), 030
doi:10.1088/1126-6708/2002/06/030
[arXiv:hep-ph/0204343 [hep-ph]].

\bibitem{Berges:2020fwq}
J.~Berges, M.~P.~Heller, A.~Mazeliauskas and R.~Venugopalan,
``QCD thermalization: Ab initio approaches and interdisciplinary connections,''
Rev. Mod. Phys. \textbf{93} (2021) no.3, 035003
doi:10.1103/RevModPhys.93.035003 [arXiv:2005.12299 [hep-th]].

\bibitem{Kurkela:2018oqw}
A.~Kurkela and A.~Mazeliauskas,
``Chemical equilibration in weakly coupled QCD,''
Phys. Rev. D \textbf{99} (2019) no.5, 054018
doi:10.1103/PhysRevD.99.054018 [arXiv:1811.03068 [hep-ph]].

\bibitem{Baier:2000sb}
R.~Baier, A.~H.~Mueller, D.~Schiff and D.~T.~Son,
``'Bottom up' thermalization in heavy ion collisions,''
Phys. Lett. B \textbf{502} (2001), 51-58
doi:10.1016/S0370-2693(01)00191-5
[arXiv:hep-ph/0009237 [hep-ph]].

\bibitem{Gribov:1972ri}
V.~N.~Gribov and L.~N.~Lipatov,
``Deep inelastic e p scattering in perturbation theory,''
Sov. J. Nucl. Phys. \textbf{15} (1972), 438.

\bibitem{Dokshitzer:1977sg}
Y.~L.~Dokshitzer,
``Calculation of the Structure Functions for Deep Inelastic Scattering and
$e^+ e^-$ Annihilation by Perturbation Theory in Quantum Chromodynamics,''
Sov. Phys. JETP \textbf{46} (1977), 641.

\bibitem{Altarelli:1977zs}
G.~Altarelli and G.~Parisi,
``Asymptotic Freedom in Parton Language,''
Nucl. Phys. B \textbf{126} (1977), 298, doi:10.1016/0550-3213(77)90384-4

\bibitem{Boyarsky:2020ani}
A.~Boyarsky, V.~Cheianov, O.~Ruchayskiy and O.~Sobol,
``Equilibration of the chiral asymmetry due to finite electron mass in
electron-positron plasma,'' Phys. Rev. D \textbf{103} (2021) no.1, 013003,
doi:10.1103/PhysRevD.103.013003 [arXiv:2008.00360 [hep-ph]].

\bibitem{Boyarsky:2020cyk}
A.~Boyarsky, V.~Cheianov, O.~Ruchayskiy and O.~Sobol,
``Evolution of the Primordial Axial Charge across Cosmic Times,''
Phys. Rev. Lett. \textbf{126} (2021) no.2, 021801,
doi:10.1103/PhysRevLett.126.021801 [arXiv:2007.13691 [hep-ph]].

\bibitem{Reya:1979zk}
E.~Reya, ``Perturbative Quantum Chromodynamics,''
Phys. Rept. \textbf{69} (1981), 195
doi:10.1016/0370-1573(81)90036-3

\bibitem{Abramowicz:1982jd}
H.~Abramowicz, \textit{et al.},
``Tests of {QCD} and Nonasymptotically Free Theories of the Strong Interaction
by an Analysis of the Nucleon Structure Functions Xf(3), F(2), and $\bar{q}$,''
Z. Phys. C \textbf{13} (1982), 199
doi:10.1007/BF01575772

\bibitem{Mazeliauskas:2018yef}
A.~Mazeliauskas and J.~Berges,
``Prescaling and far-from-equilibrium hydrodynamics in the quark-gluon plasma,''
Phys. Rev. Lett. \textbf{122} (2019) 122301, 
doi:10.1103/PhysRevLett.122.122301
[arXiv:1810.10554 [hep-ph]].

\bibitem{Bertone:2004pz}
G.~Bertone, D.~Hooper and J.~Silk,
``Particle dark matter: Evidence, candidates and constraints,''
Phys. Rept. \textbf{405} (2005), 279, 
doi:10.1016/j.physrep.2004.08.031 [arXiv:hep-ph/0404175 [hep-ph]].

\bibitem{Ghiglieri:2021vcq}
J.~Ghiglieri and M.~Laine,
``Smooth interpolation between thermal Born and LPM rates,''
[arXiv:2110.07149 [hep-ph]].

\end{thebibliography}
\end{document}